\documentclass[a4paper,11pt]{article}
\pdfoutput=1 

\usepackage{jheppub} 

\usepackage[T1]{fontenc} 

\title{$W$-Boson Mass Anomaly from a General $SU(2)_{L}$ Scalar Multiplet}

\author[a,b,c,d]{Jiajun Wu,}
\author[e,a,b,d]{Da Huang,}
\author[a,b,d]{Chao-Qiang Geng}

\affiliation[a]{School of Fundamental Physics and Mathematical Sciences, Hangzhou Institute for Advanced Study, UCAS, Hangzhou 310024, China}
\affiliation[b]{University of Chinese Academy of Sciences (UCAS), Beijing 100049, China}
\affiliation[c]{CAS Key Laboratory of Theoretical Physics, Institute of Theoretical Physics, Chinese Academy of Sciences, Beijing, 100190, China}
\affiliation[d]{International Centre for Theoretical Physics Asia-Pacific, Beijing/Hangzhou, China}
\affiliation[e]{National Astronomical Observatories, Chinese Academy of Sciences, Beijing, 100012, China}

\emailAdd{wujiajun@itp.ac.cn}	
\emailAdd{dahuang@bao.ac.cn}
\emailAdd{geng@phys.nthu.edu.tw}

\abstract{
We explain the $W$-boson mass anomaly by introducing an $SU(2)_L$ scalar multiplet with general isospin and hypercharge {in the case without its vacuum expectation value}. It is shown that the dominant contribution from the scalar multiplet to the $W$-boson mass arises at one-loop level, which can be expressed in terms of the electroweak (EW) oblique parameters $T$ and $S$ at leading order. We firstly rederive the general formulae of $T$ and $S$ induced by a scalar multiplet of EW charges, confirming the results in the literature. We then study several specific examples of great phenomenological interest by applying these general expressions. As a result, it is found that the model with a scalar multiplet in an $SU(2)_L$ real representation with $Y=0$ cannot generate 
the required $M_W$ correction since it leads to vanishing values of $T$ and $S$. On the other hand, the cases with scalars in a complex representation under $SU(2)_L$ with a general hypercharge can explain the $M_W$ excess observed by CDF-\uppercase\expandafter{\romannumeral2} due to nonzero $T$ and $S$. We further take into account of the strong constraints from the perturbativity and the EW global fit of the precision data, and vary the isospin representation and hypercharge of the additional scalar multiplet, in order to assess the extent of the model to solve the $W$-boson mass anomaly. It turns out that these constraints play  important roles in setting limits on the model parameter space. We also briefly describe the collider signatures of the extra scalar multiplet, especially when it contains long-lived heavy highly charged states.   }

\begin{document} 
\maketitle
\flushbottom

\section{Introduction}
\label{sec:intro}
The CDF-\uppercase\expandafter{\romannumeral2} Collaboration has recently reported their new measurement of the $W$-boson mass~\cite{CDF:2022hxs} as follows
\begin{equation}
	M_{W,{\rm CDF\mbox{-}\uppercase\expandafter{\romannumeral2}}}=80.4335 \pm 0.0094\enspace {\rm GeV}\,.
\end{equation}
However, the most recent prediction of $M_W$ in the Standard Model (SM) is given by~\cite{PDG},
\begin{equation}
	M_{W,{\rm SM}}=80.357 \pm 0.006\enspace {\rm GeV}\,.
\end{equation}
Thus, it is easy to see that the discrepancy between the latest CDF-\uppercase\expandafter{\romannumeral 2} value and the SM calculation of $M_W$ has reached over 7$\sigma$ confidence level (CL.). If this $M_W$ anomaly is further confirmed in the future, then it would provide us a novel hint towards  
New Physics (NP) beyond the SM (BSM). In the literature, there have already been many BSM attempts~\cite{Sakurai:2022hwh,Peli:2022ybi,Dcruz:2022dao,Asai:2022uix,Fan:2022dck,Lu:2022bgw,Song:2022xts,Bahl:2022xzi,Babu:2022pdn,Heo:2022dey,Ahn:2022xeq,Ghorbani:2022vtv,Lee:2022gyf,Abouabid:2022lpg,Benbrik:2022dja,Botella:2022rte,Kim:2022hvh,Kim:2022xuo,Appelquist:2022qgl,Benincasa:2022elt,Arhrib:2022inj,Han:2022juu,Abdallah:2022shy,Barrie:2022cub,Cheng:2022jyi,Du:2022brr,FileviezPerez:2022lxp,Kanemura:2022ahw,Mondal:2022xdy,Borah:2022obi,Addazi:2022fbj,Heeck:2022fvl,Chen:2022ocr,Evans:2022dgq,Ghosh:2022zqs,Ma:2022emu,Bahl:2022gqg,Penedo:2022gej,Cheng:2022hbo,Butterworth:2022dkt,Song:2022jns,Blennow:2022yfm,Lee:2022nqz,Cheung:2022zsb,Crivellin:2022fdf,Ghoshal:2022vzo,Kawamura:2022uft,Popov:2022ldh,Cao:2022mif,Dermisek:2022xal,Li:2022gwc,He:2022zjz,Chowdhury:2022dps,Zeng:2022llh,Zhang:2022nnh,Zeng:2022lkk,Du:2022fqv,Baek:2022agi,Cheng:2022aau,Faraggi:2022emm,Cai:2022cti,Thomas:2022gib,Wojcik:2022rtk,Afonin:2022cbi,Allanach:2022bik,Nagao:2022dgl,VanLoi:2022eir,deBlas:2022hdk,Fan:2022yly,Bagnaschi:2022whn,Paul:2022dds,Balkin:2022glu,Cirigliano:2022qdm,Almeida:2022lcs,Gupta:2022lrt,Guedes:2022cfy,Liu:2022vgo,Yang:2022gvz,Du:2022pbp,Tang:2022pxh,Athron:2022isz,Sun:2022zbq,Zheng:2022irz,Lazarides:2022spe,Yuan:2022cpw,Coy:2021hyr,DAlise:2022ypp,Strumia:2022qkt,Arias-Aragon:2022ats,Asadi:2022xiy,DiLuzio:2022xns,Gu:2022htv,Biekotter:2022abc,DiLuzio:2022ziu,Endo:2022kiw,Nagao:2022oin,Arcadi:2022dmt,Chowdhury:2022moc,Bhaskar:2022vgk,Borah:2022zim,Batra:2022org,Batra:2022pej,Wang:2022dte,Li:2022eby,Kawamura:2022fhm,Zhou:2022cql,Rizzo:2022jti,deGiorgi:2022xhr,Amoroso:2022rly,Primulando:2022vip,Pfeifer:2022yrs,Chung:2022avf,Lin:2022khg,Diaz:2022vdo,Liu:2022zie,Rodriguez:2022hsj,Domingo:2022pde,Kim:2022axk,Barger:2022wih,Benakli:2022gjn,Senjanovic:2022zwy,Bandyopadhyay:2022bgx,Kumar:2022rcf,Heckman:2022the,Frandsen:2022xsz,Basiouris:2022wei,Kim:2022zhj} to explain the $M_W$ discrepancy. Among various BSM scenarios, the extension of the SM Higgs sector by introducing an additional $SU(2)_L$ multiplet~\cite{Sakurai:2022hwh,Peli:2022ybi,Dcruz:2022dao,Asai:2022uix,Fan:2022dck,Lu:2022bgw,Song:2022xts,Bahl:2022xzi,Babu:2022pdn,Heo:2022dey,Ahn:2022xeq,Ghorbani:2022vtv,Lee:2022gyf,Abouabid:2022lpg,Benbrik:2022dja,Botella:2022rte,Kim:2022hvh,Kim:2022xuo,Appelquist:2022qgl,Benincasa:2022elt,Arhrib:2022inj,Han:2022juu,Abdallah:2022shy,Barrie:2022cub,Cheng:2022jyi,Du:2022brr,FileviezPerez:2022lxp,Kanemura:2022ahw,Mondal:2022xdy,Borah:2022obi,Addazi:2022fbj,Heeck:2022fvl,Chen:2022ocr,Evans:2022dgq,Ghosh:2022zqs,Ma:2022emu,Bahl:2022gqg,Penedo:2022gej,Cheng:2022hbo,Butterworth:2022dkt,Song:2022jns} is a promising direction. On the one hand, the modification of the scalar sector is intimately related to the true mechanism for the electroweak (EW) gauge symmetry breaking and the associated hierarchy problem, which might be probed by measuring the EW oblique parameters~\cite{Peskin:1990zt,Peskin:1991sw,Maksymyk:1993zm,Burgess:1993mg,Lavoura:1993nq,Albergaria:2021dmq} and the trilinear Higgs coupling~\cite{Degrassi:2021uik,McCullough:2013rea,Cao:2015oxx,Bizon:2016wgr,deBlas:2016ojx}. On the other hand, the introduced scalar multiplet could solve many puzzles in the SM, such as the nature of  dark matter (DM)~\cite{Cirelli:2005uq,Cirelli:2009uv,Guo:2010hq,Barbieri:2006dq,LopezHonorez:2010eeh,Gonderinger:2012rd}, the generation of the matter-anti-matter asymmetry in the Universe~\cite{Cline:2012hg,Grzadkowski:2018nbc,Cline:2021iff,Morrissey:2012db}, as well as the characteristics of the EW phase transition and its associated stochastic gravitational wave signals~\cite{Chowdhury:2011ga,Hashino:2018zsi,Chiang:2017nmu,Kannike:2019wsn,Chiang:2020yym,Chiang:2019oms,Cai:2017tmh,Chao:2017vrq,Ellis:2018mja,Alves:2018jsw,Zhou:2018zli,Bian:2019kmg,Ghosh:2020ipy,Zhou:2020irf,Lu:2022zpn,Zhou:2022mlz,Cai:2022bcf,Hashino:2018wee}. Therefore, unveiling the structure of the scalar sector could help us deeper our understanding of the overall pictures of the SM and the physics beyond it.

In the light of the potential importance of the scalar multiplet extension of the SM, we focus on its solution to the $W$-boson mass anomaly. Note that previous studies have only concentrated on the several specific models by including a scalar singlet~\cite{Sakurai:2022hwh,Peli:2022ybi,Dcruz:2022dao,Asai:2022uix}, the second Higgs doublet~\cite{Fan:2022dck,Lu:2022bgw,Song:2022xts,Bahl:2022xzi,Babu:2022pdn,Heo:2022dey,Ahn:2022xeq,Ghorbani:2022vtv,Lee:2022gyf,Abouabid:2022lpg,Benbrik:2022dja,Botella:2022rte,Kim:2022hvh,Kim:2022xuo,Appelquist:2022qgl,Benincasa:2022elt,Arhrib:2022inj,Han:2022juu,Abdallah:2022shy}, and a scalar triplet with $Y=0$ or $1$~\cite{Barrie:2022cub,Cheng:2022jyi,Du:2022brr,FileviezPerez:2022lxp,Kanemura:2022ahw,Mondal:2022xdy,Borah:2022obi,Addazi:2022fbj,Heeck:2022fvl,Chen:2022ocr,Evans:2022dgq,Ghosh:2022zqs,Ma:2022emu,Bahl:2022gqg,Penedo:2022gej,Cheng:2022hbo,Butterworth:2022dkt}, respectively. In the present work, instead of studying a particular model, we shall vary the EW $SU(2)_L$ isospin representation $J$ and the hypercharge $Y$ of the additional scalar {in the absence of its vacuum expectation value (VEV),} and see their effects on the interpretation of the CDF-\uppercase\expandafter{\romannumeral 2} $W$-boson excess. Note that when the scalar multiplet does not carry any VEV\footnote{The models by introducing a scalar multiplet with its VEV have already been studied in {\it e.g.}, Refs.~\cite{Bahl:2022gqg,Cheng:2022hbo,Asadi:2022xiy}.}, the dominant contributions to the $W$-mass correction are provided by those at one-loop level, which can be represented as the linear combination of the EW oblique parameters $T$ and $S$~\cite{Asadi:2022xiy,Gu:2022htv}. Thus, we firstly rederive the analytic formulae for $T$ and $S$ from a general scalar multiplet, confirming the results on $M_W$ in the literature. We then apply these simple expressions to several scenarios of physical interest, like a real or complex multiplet with $Y=0$, the cases with $J=Y$, and the variation of $Y$ with a fixed $J$. In our phenomenological studies, we also take into account the constraints from EW global fits and the perturbativity, which directly constrain the present scalar models.

The paper is organized as follows. In Sec.\ref{sec2}, we begin by setting the notation of a general $SU(2)_{L}$ scalar multiplet, as well as its relevant terms in the Lagrangian interacting with the SM gauge and Higgs doublet bosons. We also give the general formulae of its one-loop contribution to the $W$-boson mass, which can be expressed as the linear combinations of the oblique parameters $T$ and $S$ at  leading order. In Sec.\ref{sec3}, we study several specific models of great physical interest in the scalar-multiplet extensions of the SM, and explore their ability to explain the observed $M_W$ anomaly by varying the scalar $SU(2)_L$ representations and hypercharges, where we also take into account the constraints from the perturbativity and EW global fits. We conclude in Sec.~\ref{sec4}, where we also briefly discuss possible collider signatures of the additional scalars, especially when the dimension of scalar multiplet is high. In addition, we include several appendices. In Appendix~\ref{app1}, we give the relevant part of the Lagrangian and associated Feynman rules involving the scalar multiplet. In Appendix~\ref{app2}, we show that two particular terms in the scalar potential can be represented as linear combinations of other existing terms, making them be ignored in the following discussion. In Appendix~\ref{app3}, we define several functions relating to the one-loop contributions of the scalar multiplet to the oblique parameters $T$ and $S$. In Appendix~\ref{app4}, we give the calculation details when we rederive the one-loop expressions of the $T$ and $S$ parameters from a scalar multiplet with general EW charges.

\section{General Discussions}
\label{sec2}

\subsection{The Model}
\label{sec2.1}

Generally, we label the scalar multiplet as $\Phi_{JY}$,
where $J$ denotes the weak isospin $SU(2)_L$ representation and $Y$ is the hypercharge, so that the dimension of the multiplet is $N=2J+1$. We also label the components in the scalar multiplet $\Phi_{JY}$ to be $\Phi_{I}^{Q}$ as follows:
\begin{equation}\label{ComponentPhi}
	\Phi_{JY}=\left(\begin{array}{c}
		...\\
		\varPhi_{I}^{Q}\\
		...
	\end{array}\right),
\end{equation}
where $I=J,J-1,J-2,......,-J+1,-J$ and the electric charge $Q=I+Y$. 

The EW covariant derivative after the EW symmetry breaking can be written as
\begin{equation}
	D_{\mu}=\partial_{\mu}+ieQA_{\mu}+i\frac{g}{c_{W}}\left(T_{3}-Qs_{W}^{2}\right)Z_{\mu}+ig\left(W_{\mu}^{+}T_{+}+W_{\mu}^{-}T_{-}\right),
\end{equation}
where $T_{k}~(k=1,2,3)$ are the three generators of the $SU(2)_L$ group, and the ladder operator $T_{\pm}$ are given by $T_{\pm}=(T_{1} \pm iT_{2})/\sqrt{2}$. The eigenvalues of $T_{3}$ can be labeled by $I$ as in Eq.~(\ref{ComponentPhi}). The actions of $T_{3}$ and $T_{\pm}$ on $\Phi_{JY}$ are given by
\begin{equation}
	T_{3}\Phi_{JY}=T_{3}\left(\begin{array}{c}
		...\\
		\varPhi_{I}^{Q}\\
		...
	\end{array}\right)=\left(\begin{array}{c}
		...\\
		I\varPhi_{I}^{Q}\\
		...
	\end{array}\right),
\end{equation}
\begin{equation}
	T_{+}\Phi_{JY}=T_{+}\left(\begin{array}{c}
		...\\
		\varPhi_{I}^{Q}\\
		...
	\end{array}\right)=\left(\begin{array}{c}
		...\\
		N_{I}\varPhi_{I-1}^{Q}\\
		...
	\end{array}\right),
\end{equation}
\begin{equation}\label{TMinus}
	T_{-}\Phi_{JY}=T_{-}\left(\begin{array}{c}
		...\\
		\varPhi_{I}^{Q}\\
		...
	\end{array}\right)=\left(\begin{array}{c}
		...\\
		N_{I+1}\varPhi_{I+1}^{Q}\\
		...
	\end{array}\right),
\end{equation}
where $N_{I}=\sqrt{{(J+I)(J-I+1)}/{2}}$ and we have only shown the element with the quantum number of $T_3$ equal to $I$ in the multiplet. 
Then $D^{\mu}\Phi_{JY}$ can be written as
\begin{equation}
	\begin{aligned}
		D^{\mu}\Phi_{JY}= &\partial^{\mu}\varPhi_{I}^{Q}+ieQA_{\mu}\varPhi_{I}^{Q}+i\frac{g}{c_{W}}\left(I-Qs_{W}^{2}\right)Z_{\mu}\varPhi_{I}^{Q}\\
		&+igW_{\mu}^{+}N_{I}\varPhi_{I-1}^{Q}+igW_{\mu}^{-}N_{I+1}\varPhi_{I+1}^{Q}\,.
	\end{aligned}
\end{equation}
The gauge interaction terms stemming from the kinetic term $\left(D_{\mu}\Phi_{JY}\right)^{\dagger}D^{\mu}\Phi_{JY}$ are shown explicitly in Appendix~\ref{app1}.

In this kind of models, the potential constructed by the Higgs doublet $H$ and the scalar multiplet $\Phi_{JY}$ is given by
\begin{equation}\label{potential}
	\begin{aligned}
		V\left(H,\Phi_{JY}\right)=&-\mu_{H}^{2}H^{\dagger}H+\lambda_{H}\left(H^{\dagger}H\right)^{2}+\mu_{\Phi_{JY}}^{2}\Phi_{JY}^{\dagger}\Phi_{JY}+\lambda_{1}\left(\Phi_{JY}^{\dagger}\Phi_{JY}\right)^{2}\\
		&+\lambda_{2}\left(\Phi_{JY}^{\dagger}T_{\Phi}^{a}\Phi_{JY}\right)^{2}+\lambda_{3}\left(\Phi_{JY}^{\dagger}\Phi_{JY}\right)\left(H^{\dagger}H\right)\\
		&+\lambda_{4}\left(\Phi_{JY}^{\dagger}T_{\Phi_{JY}}^{a}\Phi_{JY}\right)\left(H^{\dagger}T_{H}^{a}H\right)+\lambda_{5}\left(\Phi_{JY}^{\dagger}T_{\Phi_{JY}}^{a}T_{\Phi_{JY}}^{b}\Phi_{JY}\right)^{2}\,.
	\end{aligned}
\end{equation}
Here, we only list the most general interaction terms for any scalar multiplet $\Phi_{JY}$. Note that the potential in Eq.~(\ref{potential}) respects an $Z_2$ symmetry. 
When the dimension of the scalar multiplet is high enough, there are not any other terms at the renormalizable level, so that this $Z_2$ symmetry becomes an accidental one at low energies. If the electrically neutral component is the lightest one in the multiplet, then it could provide us a viable DM candidate, which could not decay due to the protection by this $Z_2$ symmetry. Such a scenario is the so-called minimal DM~\cite{Cirelli:2005uq}. However, for some specific $SU(2)_L \times U(1)_Y$ representations with low dimensions, there can be additional $Z_2$-symmetry breaking terms. For example, in the Type-II seesaw model~\cite{Konetschny:1977bn,Cheng:1980qt,Magg:1980ut,Schechter:1980gr,Lazarides:1980nt,Mohapatra:1980yp}, a weak-isospin triplet with $Y=1$, $\Phi_{1,1}$, can give rise to the $Z_2$-odd interaction as $H^\dagger \Phi_{1,1} \tilde{H}+{\rm h.c.}$, where $\tilde{H} \equiv i\sigma_2 H^*$ with $H^*$ denoting the complex conjugate of the Higgs doublet $H$. Nevertheless, such $Z_2$-odd terms are not relevant to our study on the one-loop corrections to the $W$-boson mass, so that we shall ignore them in the following. 
In addition, we can also write down other two operators for any scalar multiplet:
\begin{equation}\label{O67}
 O_6 + O_7 \equiv 	\lambda_{6}\left(H^{\dagger}\Phi_{JY}\right)\left(\Phi_{JY}^{\dagger}H\right)+\lambda_{7}\left|\widetilde{H}\Phi_{JY}\right|^{2}\,.
\end{equation}
However, as shown in Appendix~\ref{app2}, we find that they are not independent, as they can be represented as linear combinations of terms proportional to $\lambda_{3}$ and $\lambda_{4}$. Thus, neither of them can induce NP effects, and would be discussed late on.  

\subsection{Oblique Parameters and  the $W$-Boson Mass}
\label{sec2.2}
NP effects in the EW sector are usually imprinted by three oblique parameters, $T$, $S$, and $U$~\cite{Peskin:1991sw,Peskin:1990zt}. 
In particular, the one-loop correction to the $W$-boson mass can be expressed as follows~\cite{Peskin:1991sw,Maksymyk:1993zm,Burgess:1993mg}
\begin{equation}\label{wmass}
	M_{W}=M_{W,SM}\left(1-\frac{\alpha\left(M_{Z}^{2}\right)}{4\left(c_{W}^{2}-s_{W}^{2}\right)}\left(S-2c_{W}^{2}T\right)+\frac{\alpha\left(M_{Z}^{2}\right)}{8s_{W}^{2}}U\right)\,,
\end{equation}
where $\alpha\equiv e^{2}/(4\pi)=g^{2}s_{W}^{2}/(4\pi)$ is the fine-structure constant, $s_{W}\equiv  \sin\theta_{W}$, and $c_{W}\equiv\cos\theta_{W}$, with $\theta_{W}$ denoting the Weinberg angle. Here, the oblique parameters $T$, $S$ and $U$ at one-loop level are defined as follows\footnote{{Throughout this paper, we use the definition of the oblique parameters $T$, $S$ and $U$ as in Ref.~\cite{Asadi:2022xiy}. However, according to Refs.~\cite{Lavoura:1993nq,Grimus:2007if,Cheng:2022hbo}, there are extra contributions to the oblique parameters $S$ and $U$, which arises due to the nonzero $Z$-boson mass $m_Z$. Note that these extra $S$ and $U$ corrections would be suppressed by $m^2_Z/m^2_{\Phi_I^Q}$ when the added scalar masses $m_{\Phi_I^Q}$ are large. For example, when all scalars are heavier than 300 GeV, the differences of $S$ and $U$ caused by these corrections are very small and can be safely ignored.}}
\begin{eqnarray}
	S&\equiv&\frac{4s_{W}^{2}c_{W}^{2}}{\alpha}\left[A_{ZZ}^{\prime}\left(0\right)-\frac{c_{W}^{2}-s_{W}^{2}}{c_{W}s_{W}}A_{Z\gamma}^{\prime}\left(0\right)-A_{\gamma\gamma}^{\prime}\left(0\right)\right]\,, \nonumber\\
	T&\equiv&\frac{1}{\alpha m_{Z}^{2}}\left[\frac{A_{WW}\left(0\right)}{c_{W}^{2}}-A_{ZZ}\left(0\right)\right]\,, \nonumber\\
	U&\equiv&\frac{4s_{W}^{2}}{\alpha}\left[A_{WW}^{\prime}\left(0\right)-\frac{c_{W}}{s_{W}}A_{Z\gamma}^{\prime}\left(0\right)-A_{\gamma\gamma}^{\prime}\left(0\right)\right]-S\,,
\end{eqnarray}
where the functions $A^{(\prime)}_{VV^\prime} (q^2)$ can be defined in terms of the vacuum polarization tensors for the EW gauge bosons $V^{(\prime)}$ as follows
\begin{equation}\label{DefAA}
	\Pi_{VV^{'}}^{\mu\nu}\left(q\right)=g^{\mu\nu}A_{VV^{'}}\left(q^{2}\right)+q^{\mu}q^{\nu}A^\prime_{VV^{'}}\left(q^{2}\right)\,,
\end{equation}
with $q$ as the four-momentum of the corresponding vector bosons. However, as shown in Ref.~\cite{Asadi:2022xiy}, $T$ and $S$ are induced by dimension-6 operators, while $U$ by a dimension-8 one so that it is suppressed significantly.  
Therefore, it is expected that the dominant one-loop contribution to the $W$-boson mass is given by $T$ and $S$ for a NP model extended by a scalar multiplet, whereas the effect from $U$ can be ignored. {We have explicitly checked that, when the added scalar masses are all larger than 300~GeV and their multiplet dimension is confined to be smaller than 10, the contribution of $U$ to the $W$ mass is always smaller than the leading ones from $T$ and/or $S$ by at least one order of magnitude, which confirms the above expectation. We shall come back to this issue in the last section.}

\begin{figure}[ht]
	\begin{center}
		\includegraphics[width= 0.95 \linewidth]{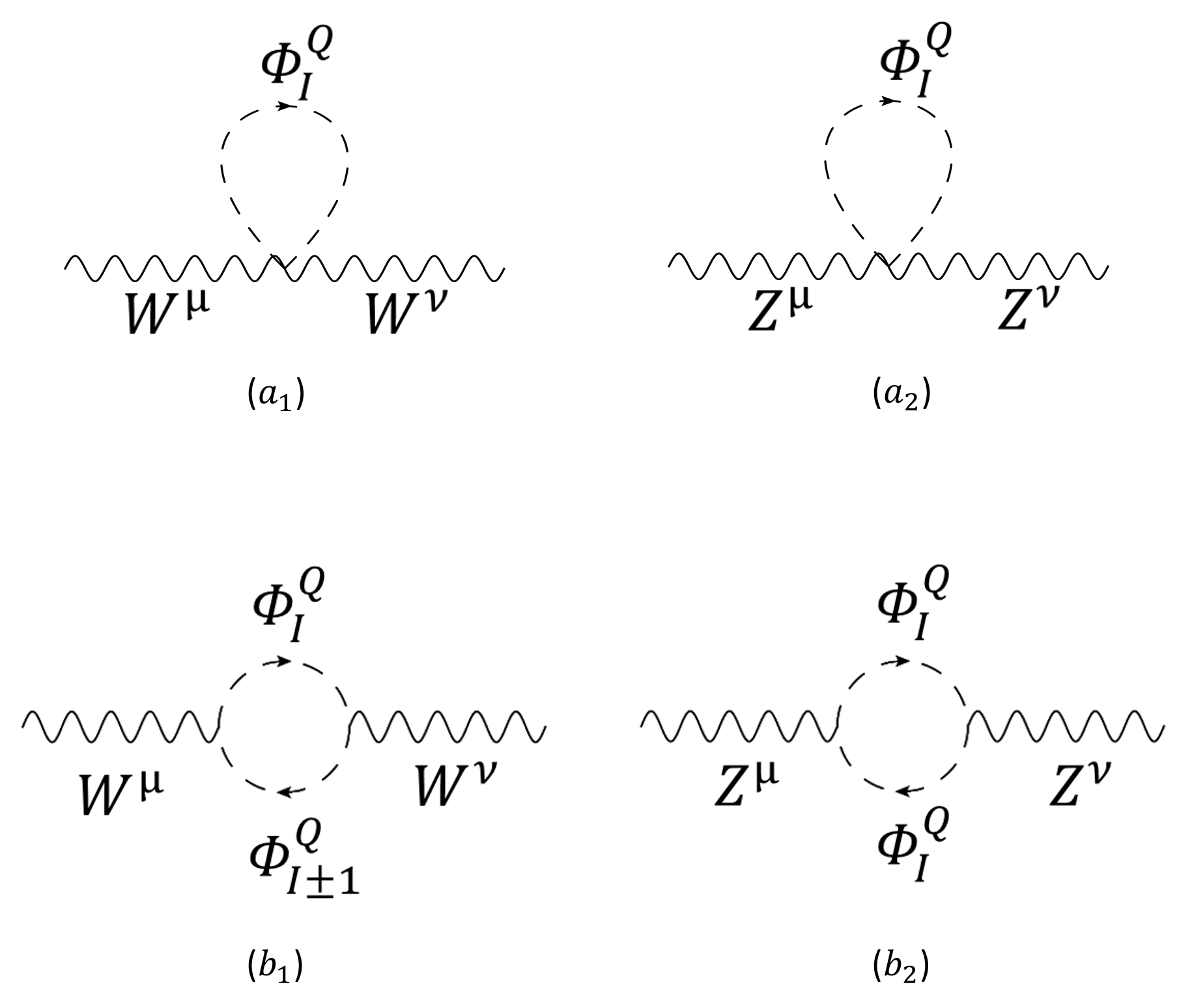}
	\end{center}
	\caption{One-loop Feynman diagrams for the $WW$ and $ZZ$ vacuum polarization tensors, which could give the leading-order  contribution to the $T$ parameter. Diagrams $(a_{1,2})$ involve only one internal scalar line while the loops in diagrams $(b_{1,2})$ are enclosed by two scalar lines.}\label{fig1}
\end{figure}
\begin{figure}[h]
	\begin{center}
		\includegraphics[width =0.95 \linewidth]{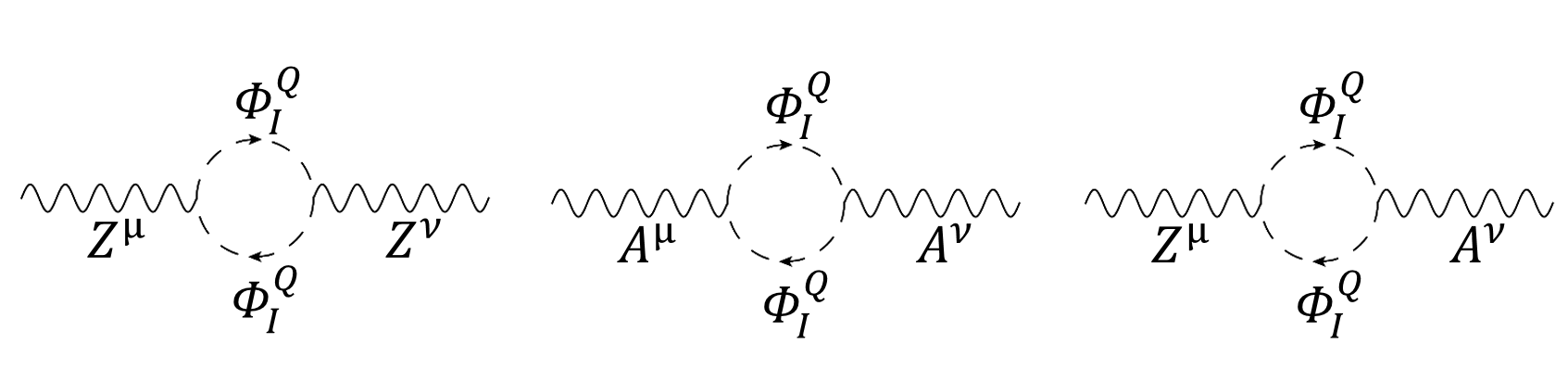}
	\end{center}
	\caption{One-loop Feynman diagrams for the $ZZ$, $AA$, and $ZA$ vacuum polarization tensors that contribute to the $S$ parameter. 
	}\label{fig2}
\end{figure}

In the present model, the scalar multiplet would induce additional contributions to the $W$-boson mass by correcting the parameters $T$ and $S$. As mentioned early, the associated corrections to these oblique parameters can be obtained by calculating the relevant Feynman diagrams shown in Figs.~\ref{fig1} and \ref{fig2} for various vacuum polarization tensors of EW gauge bosons. 
As a result, the contribution of the scalar multiplet $\Phi_{JY}$ to $T$ is given by
\begin{equation}\label{T}
	T_{\Phi_{JY}}=\frac{1}{4\pi s_{w}^{2}m_{W}^{2}}\sum_{I=-J}^{J-1}N_{I+1}^{2}F\left(m_{\varPhi_{I}^{Q}}^{2},m_{\varPhi_{I+1}^{Q}}^{2}\right)\,,
\end{equation}
where the function $F(A,B)$ is defined in Eq.~(\ref{Ffunc}) in Appendix~\ref{app3},
while the correction of $\Phi_{JY}$ to $S$ is provided as follows
\begin{equation}\label{S}
	S_{\Phi_{JY}}=-\frac{Y}{3\pi}\sum_{I=-J}^{J}I\ln m_{\varPhi_{I}^{Q}}^{2}\,.
\end{equation}
In our present work, we rederive the above scalar contributions to $T$ and $S$ independently with the details given in Appendix~\ref{app4}, where we have used the Feynman integrals and functions listed in Appendix~\ref{app3}. The final expressions confirm the results in the literature~\cite{Lavoura:1993nq,Grimus:2007if,Grimus:2008nb}.


According to Eqs.(\ref{T}) and (\ref{S}), mass differences among components in the additional scalar multiplet $\Phi_{JY}$ is necessary to generate nonzero contributions to the oblique parameters $T$ and $S$, which is required to explain the $W$-mass anomaly. Note that the mass splitting can only generated in the scalar potential of Eq.~(\ref{potential}) by the following term:
\begin{eqnarray}\label{O4}
	O_{4}=\lambda_{4}\left(\Phi_{JY}^{\dagger}T_{\Phi_{JY}}^{a}\Phi_{JY}\right)\left(H^{\dagger}T_{H}^{a}H\right)\,.
\end{eqnarray}
Consequently, we shall focus on this term in our following discussions of specific models. Furthermore, note that $O_4$ is also constrained by the perturbativity as $|\lambda_4| < 4\pi$ as a dimensionless coupling~\cite{Nebot:2007bc}. In our following discussion, we take $|\lambda_4| \leq 10$ as the perturbative limit.

\section{Specific Examples}
\label{sec3}
Having the general expression of the 1-loop correction to the $W$-boson mass in Eq.~(\ref{wmass}) and the associated formulae for the oblique parameters $T$ and $S$ in Eqs.~(\ref{T}) and (\ref{S}) at hand, we shall study several specific models of phenomenological interest, which are classified by the $SU(2)_L$ representations $J$ and the hypercharges $Y$ of the introduced scalar $\Phi_{JY}$.

\subsection{Real Multiplet with $Y=0$}
\label{sec3.1}
Firstly, we consider the  scalar multiplet to be in a real representation under the weak isospin $SU(2)_L$ group with $Y=0$. 
According to the definition, a real $SU(2)_{L}$ multiplet is related to its complex conjugate as follows
\begin{equation}\label{ComplexConj}
	\epsilon_{mm^{'}}...\epsilon_{cc^{'}}\epsilon_{bb^{'}}\epsilon_{aa^{'}}\left(\Phi^{*}\right)^{a^{'}b^{'}c^{'}...m^{'}}=\Phi_{abc...m}\,,
\end{equation}
where the Latin indices with their values to be 0 or 1 denote ones under the $SU(2)_L$ fundamental representation and
\begin{equation}
    \epsilon_{ab}=\left(\begin{array}{cc}
		0 & -1\\
		1 & 0
	\end{array}\right).
\end{equation}
We can also write the multiplet in terms of its components as
\begin{equation}\label{rpe}
	\Phi_{JY}=\frac{1}{\sqrt{2}}\left(\begin{array}{c}
		......\\
		\varPhi_{I}^{Q}\\
		......
	\end{array}\right)\,.
\end{equation}
Thus, the transformation in Eq.~(\ref{ComplexConj}) can be expressed by
\begin{equation}\label{real}
	\left(\varPhi_{I}^{Q}\right)^{*}=\varPhi_{-I}^{Q}\,.
\end{equation}

After the EW symmetry breaking, the SM Higgs doublet obtains its VEV and can be written in the unitary gauge as
\begin{equation}
	H=\left(\begin{array}{c}
		0\\
		\frac{v+h}{\sqrt{2}}
	\end{array}\right)\,.
\end{equation}
Then, we have 
\begin{equation}
	\begin{aligned}\label{lambda4}
		O_{4}=&\lambda_{4}\left(\Phi_{JY}^{\dagger}T_{\Phi_{JY}}^{a}\Phi_{JY}\right)\left(H^{\dagger}T_{H}^{a}H\right)\\
		=&\lambda_{4}\left(\Phi_{JY}^{\dagger}T_{\Phi_{JY}}^{+}\Phi_{JY}\right)\left(H^{\dagger}T_{H}^{-}H\right)+\lambda_{4}\left(\Phi_{JY}^{\dagger}T_{\Phi_{JY}}^{-}\Phi_{JY}\right)\left(H^{\dagger}T_{H}^{+}H\right)\\
		+&\lambda_{4}\left(\Phi_{JY}^{\dagger}T_{\Phi_{JY}}^{3}\Phi_{JY}\right)\left(H^{\dagger}T_{H}^{3}H\right)\\
		=&-\frac{\lambda_{4}}{4}\left(h+v\right)^{2}\sum_{I=-J}^{J}I\varPhi_{I}^{Q}\left(\varPhi_{I}^{Q}\right)^{*}\\
		\supset&-\frac{\lambda_{4}}{4}v^{2}\sum_{I=-J}^{J}I\varPhi_{I}^{Q}\left(\varPhi_{I}^{Q}\right)^{*}\,,
	\end{aligned}
\end{equation}	
where the term on the right-hand side of the last relation gives rise to the mass splitting among scalar components from $O_4$.
However, according to Eq.~(\ref{real}), we have
\begin{equation}\label{equ}
	\varPhi_{I}^{Q}\left(\varPhi_{I}^{Q}\right)^{*}=\varPhi_{-I}^{Q}\left(\varPhi_{-I}^{Q}\right)^{*},
\end{equation}
which leads to the following relations
\begin{align}\label{VanishingReal}
	\begin{cases}
		-\frac{\lambda_{4}}{4}v^{2}\left[I\varPhi_{I}^{Q}\left(\varPhi_{I}^{Q}\right)^{*}-I\varPhi_{-I}^{Q}\left(\varPhi_{-I}^{Q}\right)^{*}\right]=0 & I\textgreater 0,\\
		-\frac{\lambda_{4}}{4}v^{2}I\varPhi_{I}^{Q}\left(\varPhi_{I}^{Q}\right)^{*}=0 & I=0\,.
	\end{cases}
\end{align}
According to Eq.~(\ref{VanishingReal}), for any integer $J$, all the mass terms for $I\neq0$ are cancelled out while the $I=0$ mass term vanishes. Consequently, the $O_{4}$ term does not contribute to the mass splittings among scalars in the real multiplet. In the light of Eqs.~(\ref{T}) and (\ref{S}), this means that the model with a real multiplet cannot account for the $W$-boson mass anomaly {at the one-loop level} due to the vanishing values of $T$ and $S$. 

Note that the case with a real scalar multiplet without its VEV is usually regarded as a natural DM candidate~\cite{Cirelli:2005uq,Cirelli:2009uv} since the neutral component can be the lightest one due to the one-loop mass corrections. The result shows  that the minimal scalar DM cannot provide us with a viable solution to the $W$-mass anomaly. 

\subsection{Complex Multiplet with $Y=0$}
\label{sec3.2}
For a complex multiplet, it is seen from Eq.~(\ref{lambda4}) that 
each component in the scalar multiplet can obtain the following mass correction from $O_{4}$
\begin{equation}\label{DMY0}
	-\frac{\lambda_{4}}{4}v^{2}\sum_{I=-J}^{J}I\varPhi_{I}^{Q}\left(\varPhi_{I}^{Q}\right)^{*}\,.
\end{equation} 
Since $\Phi_{JY}$ is complex, we have
\begin{equation}
	\left(\varPhi_{I}^{Q}\right)^{*}\neq\varPhi_{-I}^{Q}\,,
\end{equation}
so that $O_4$ would induce an equal mass splitting between the adjacent components of $\Phi_{JY}$. Consequently, a complex scalar multiplet will make contribution to the $W$-boson mass. 

In the light of Eq.(\ref{S}), the scalar multiplet $\Phi_{JY}$ does not contribute to $S$ in the case of $Y=0$. Thus, the $W$-boson mass is expressed only by $T$ as follows
\begin{equation}\label{MWY0}
	M_{W}=M_{W,SM}\left(1+\frac{\alpha\left(M_{Z}^{2}\right)c_{W}^{2}}{2\left(c_{W}^{2}-s_{W}^{2}\right)}T\right).
\end{equation}
\begin{figure}[h]
	\begin{center}
		\includegraphics[width=0.48\linewidth]{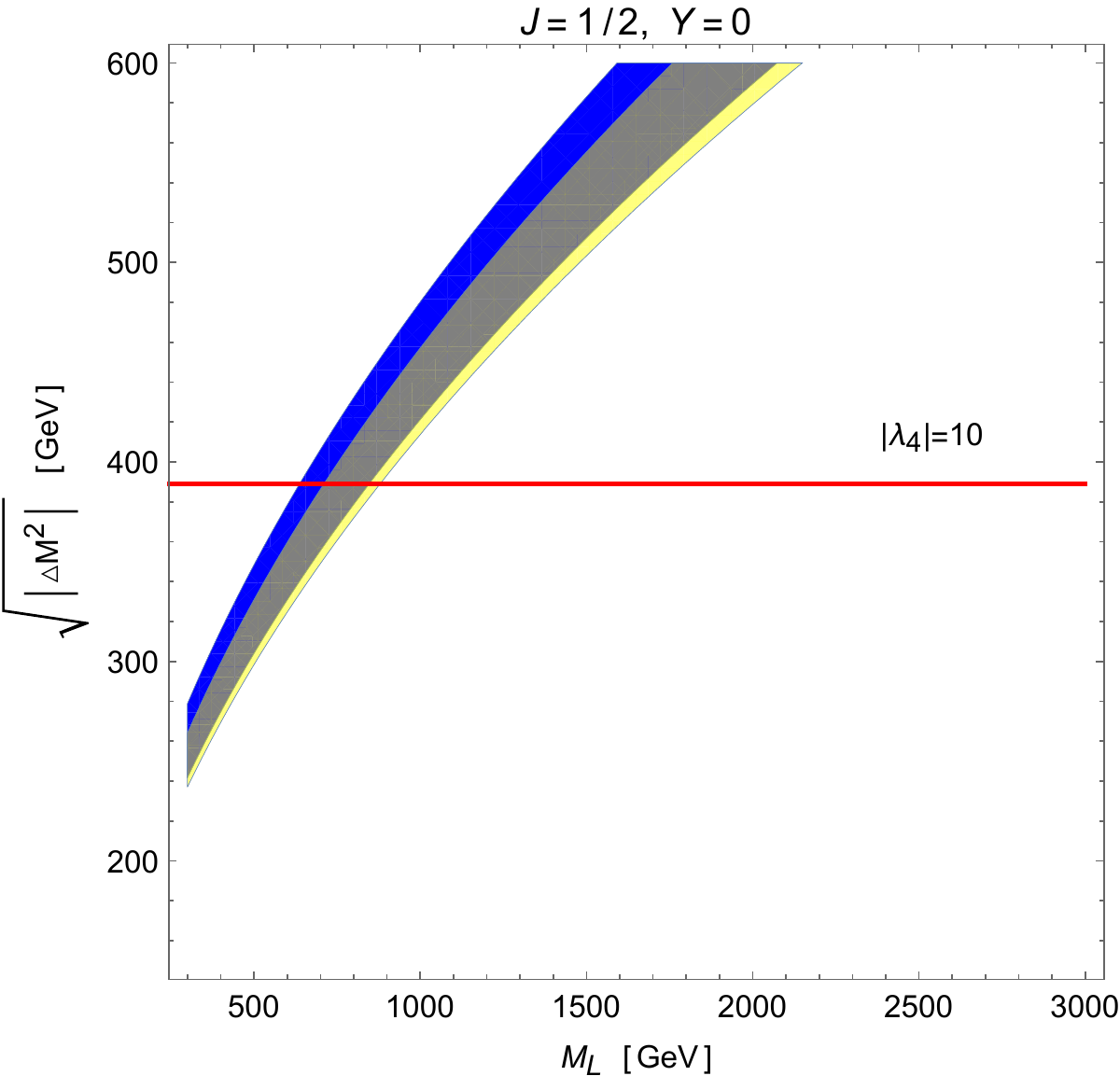}
		\includegraphics[width=0.48\linewidth]{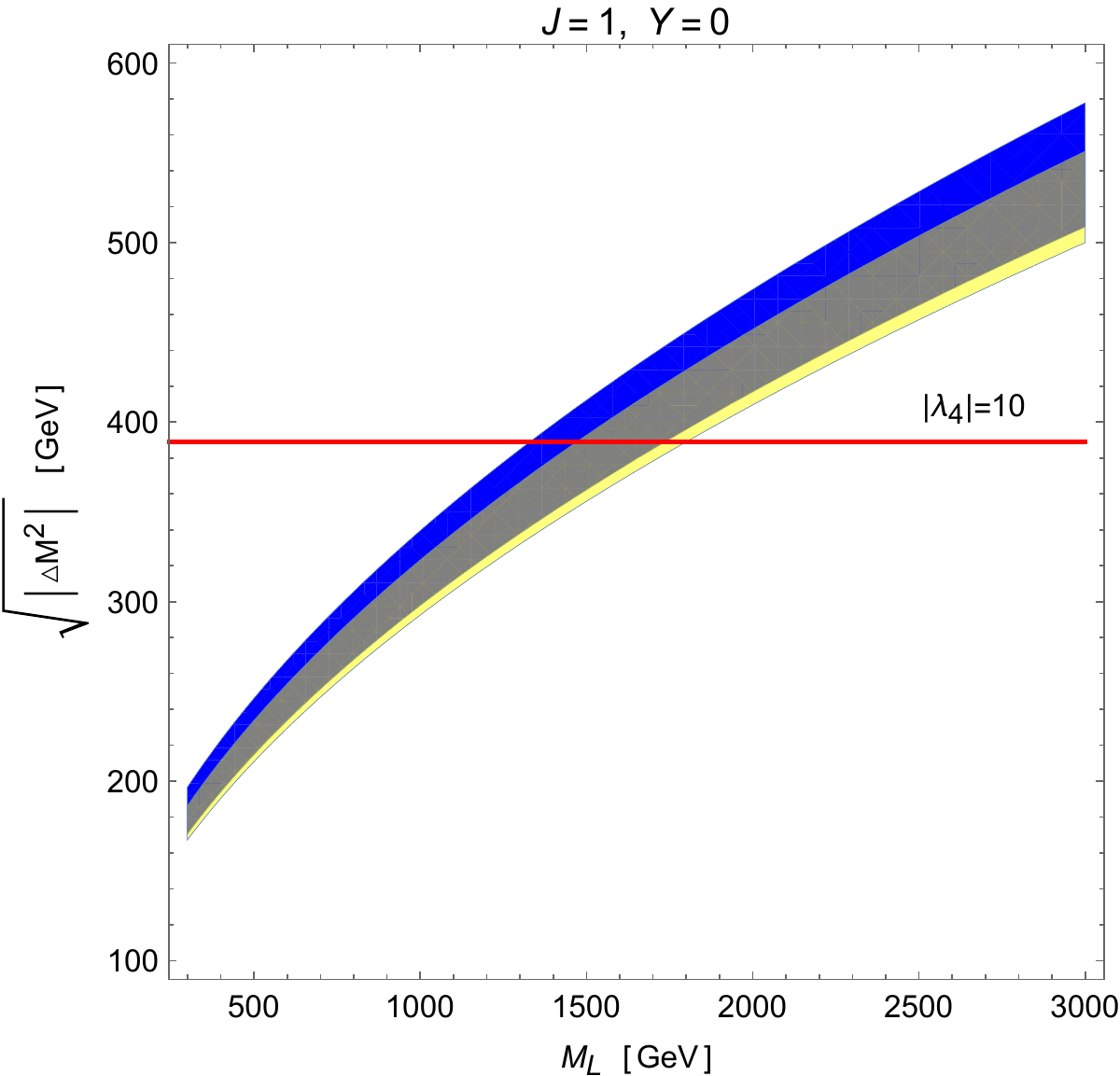}
		\includegraphics[width=0.48\linewidth]{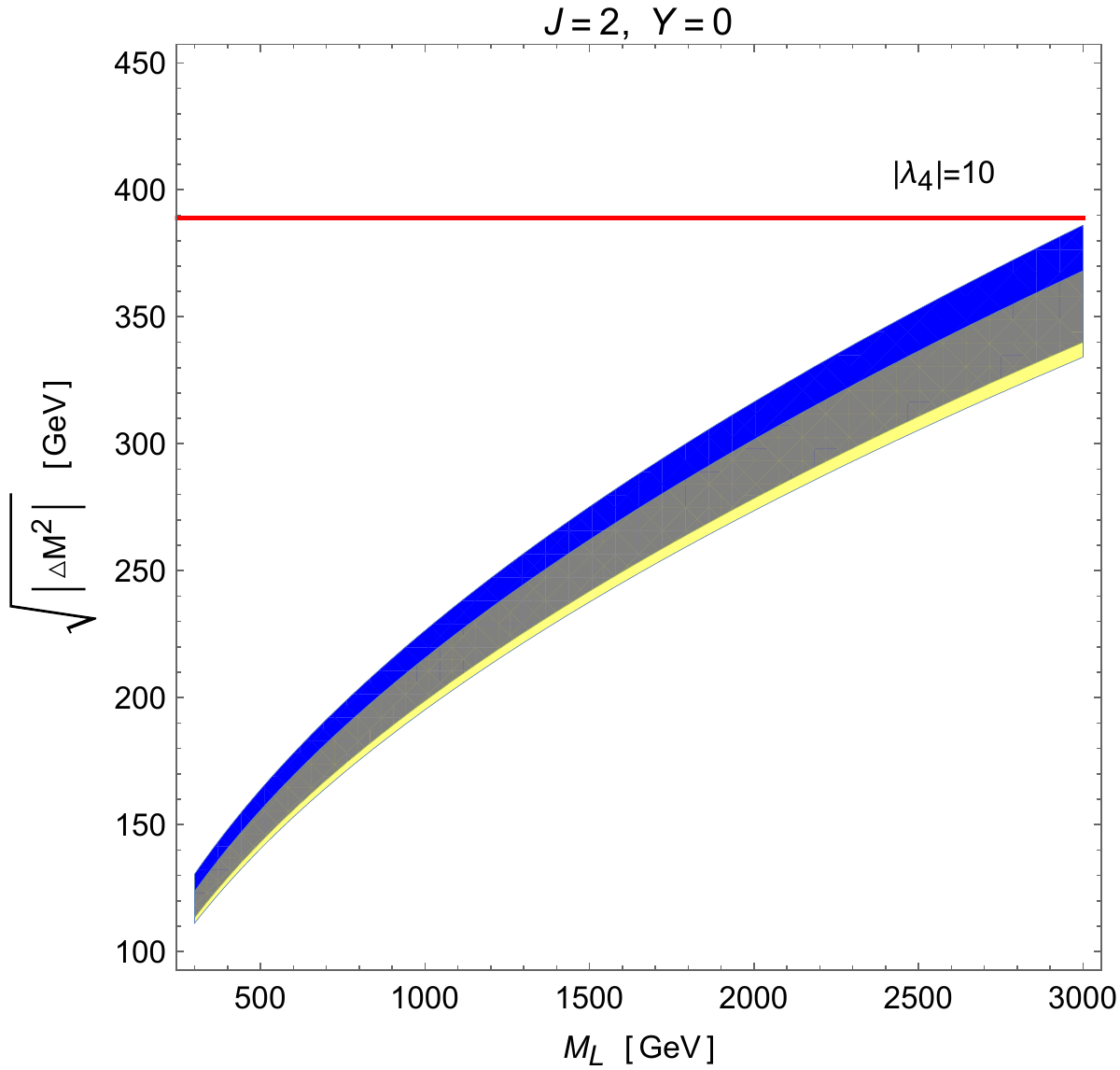}
		\includegraphics[width=0.48\linewidth]{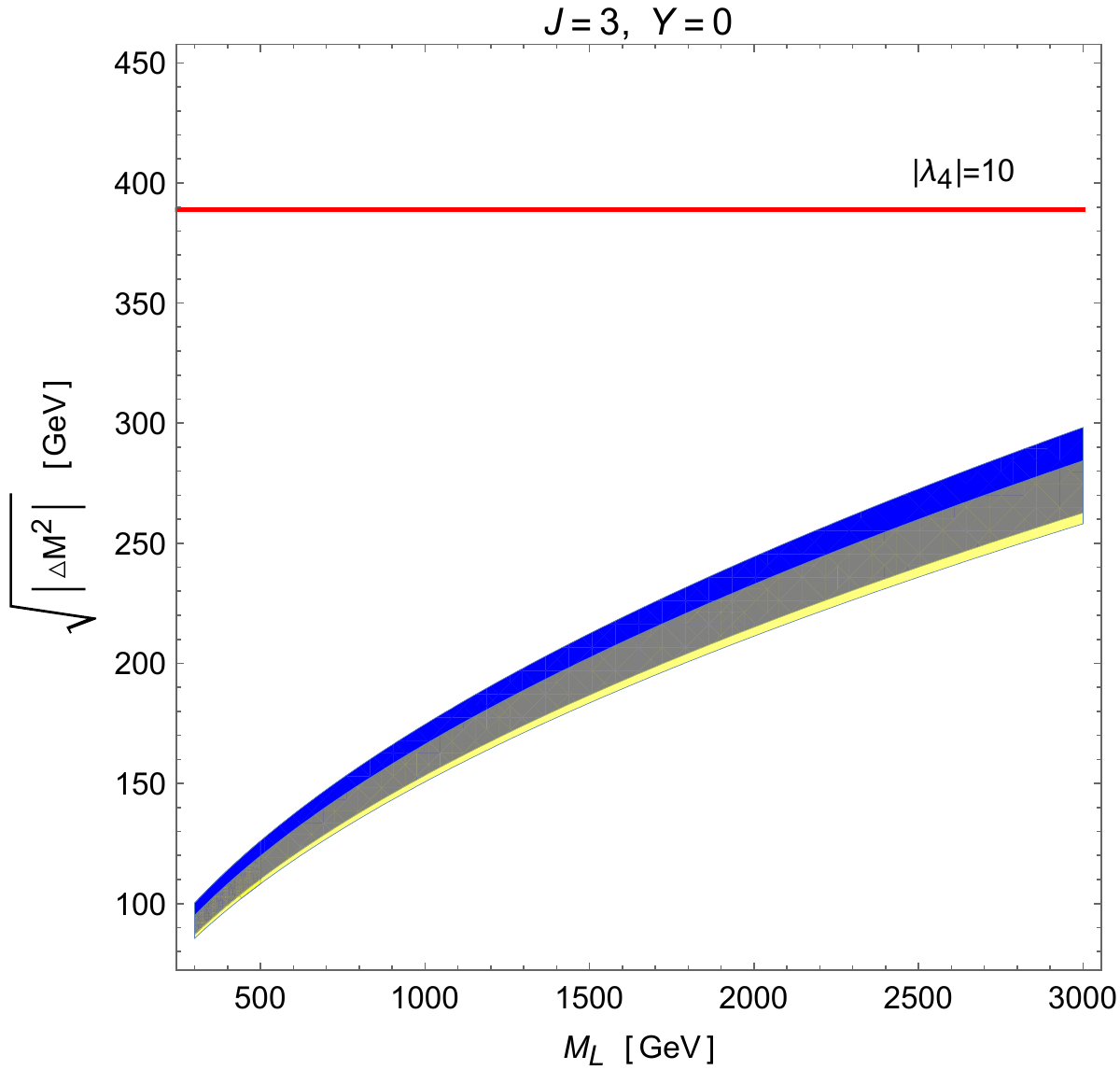}
	\end{center}
	\caption{The parameter spaces in the $M_{L}$-$\sqrt{|\Delta M^{2}|}$ plane for models with complex scalar multiplets with $Y=0$ and $J=$ 1/2, 1, 2, 3. The horizontal axis represents the mass of the lightest particle in the multiplet ($M_{L}$) {in the range from 300 GeV to 3000 GeV}, and the vertical axis represents the mass difference between adjacent components. The yellow parts in the figure are the parameter space allowed by electroweak global fit for $T$ in the $2 \sigma $ C.L. range when $S=U=0$, which is obtained from electroweak global fit in Ref.\cite{Cheng:2022hbo}. The solid blue area in the figure are the parameter space which can explain the W mass measured by ${\rm CDF\mbox{-}\uppercase\expandafter{\romannumeral2}}$ in the $2 \sigma$ C.L. range, and the yellow shadow areas are the parameter space meeting the requirements. The red solid line is the corresponding mass difference with $| \lambda_{4} |=10 $, and the areas below the red line are consistent with perturbativity.}\label{figY01}
\end{figure}
\begin{figure}[h]
	\begin{center}
		\includegraphics[width=0.48\linewidth]{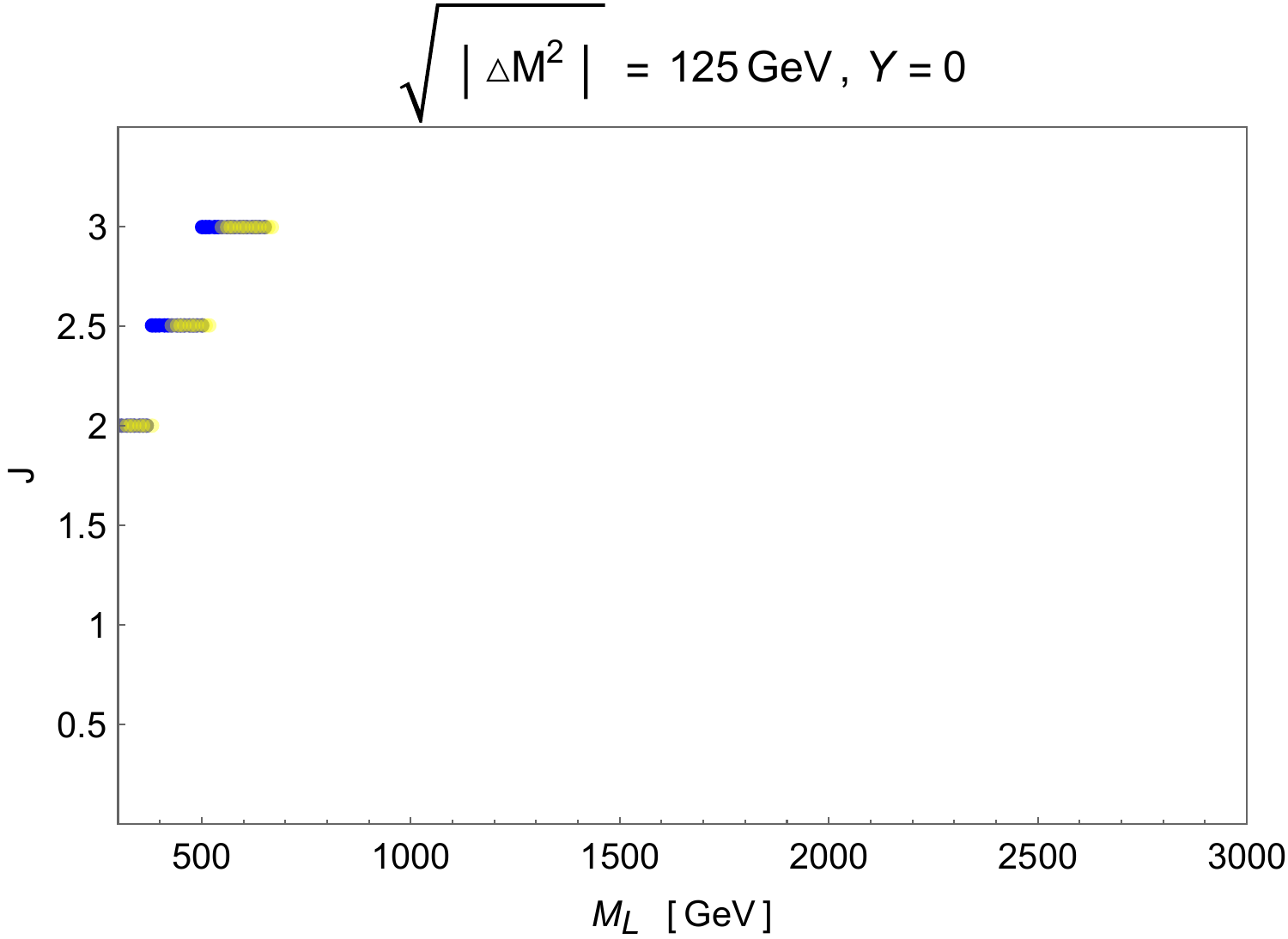}
		\includegraphics[width=0.48\linewidth]{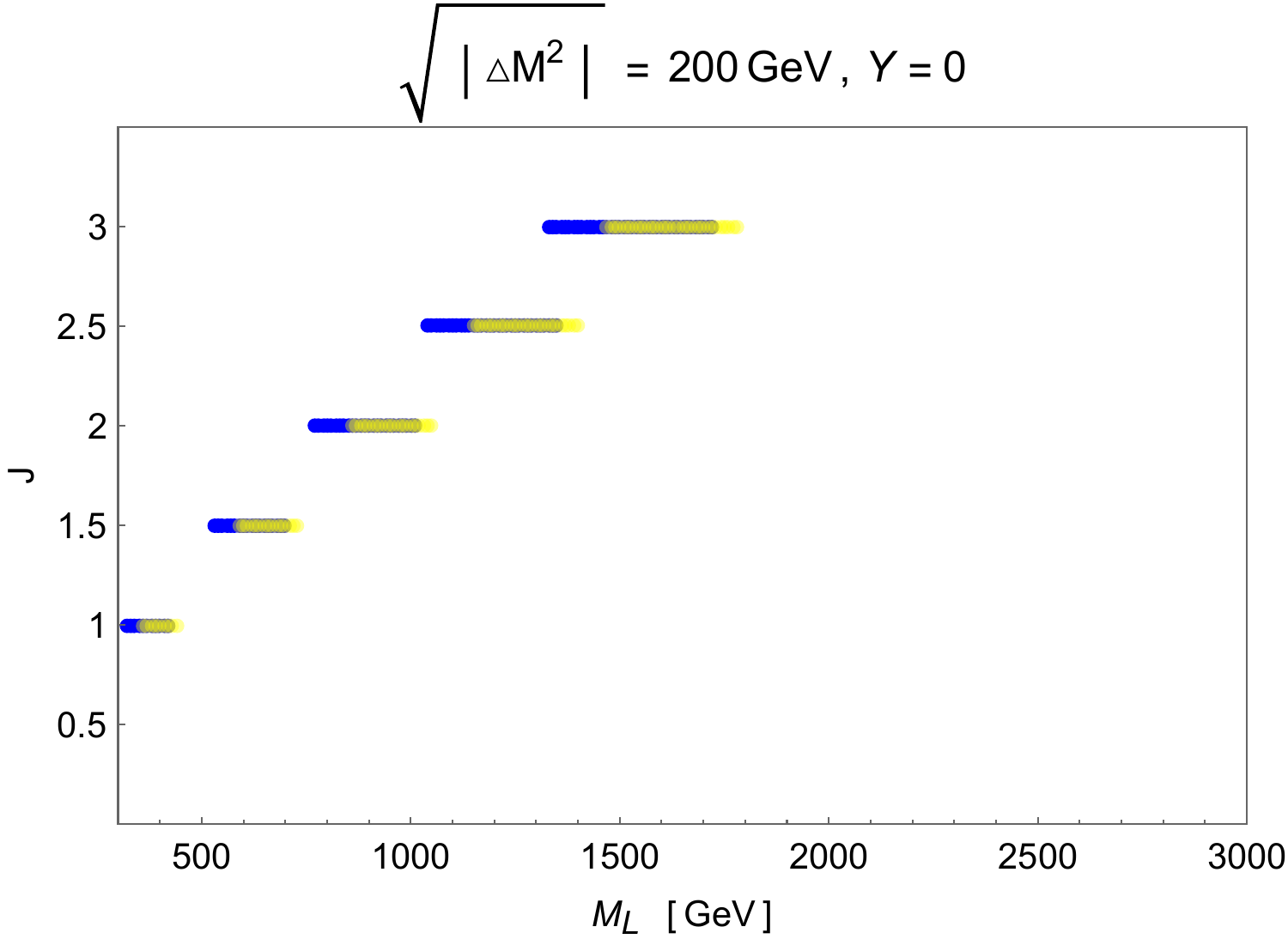}
		\includegraphics[width=0.48\linewidth]{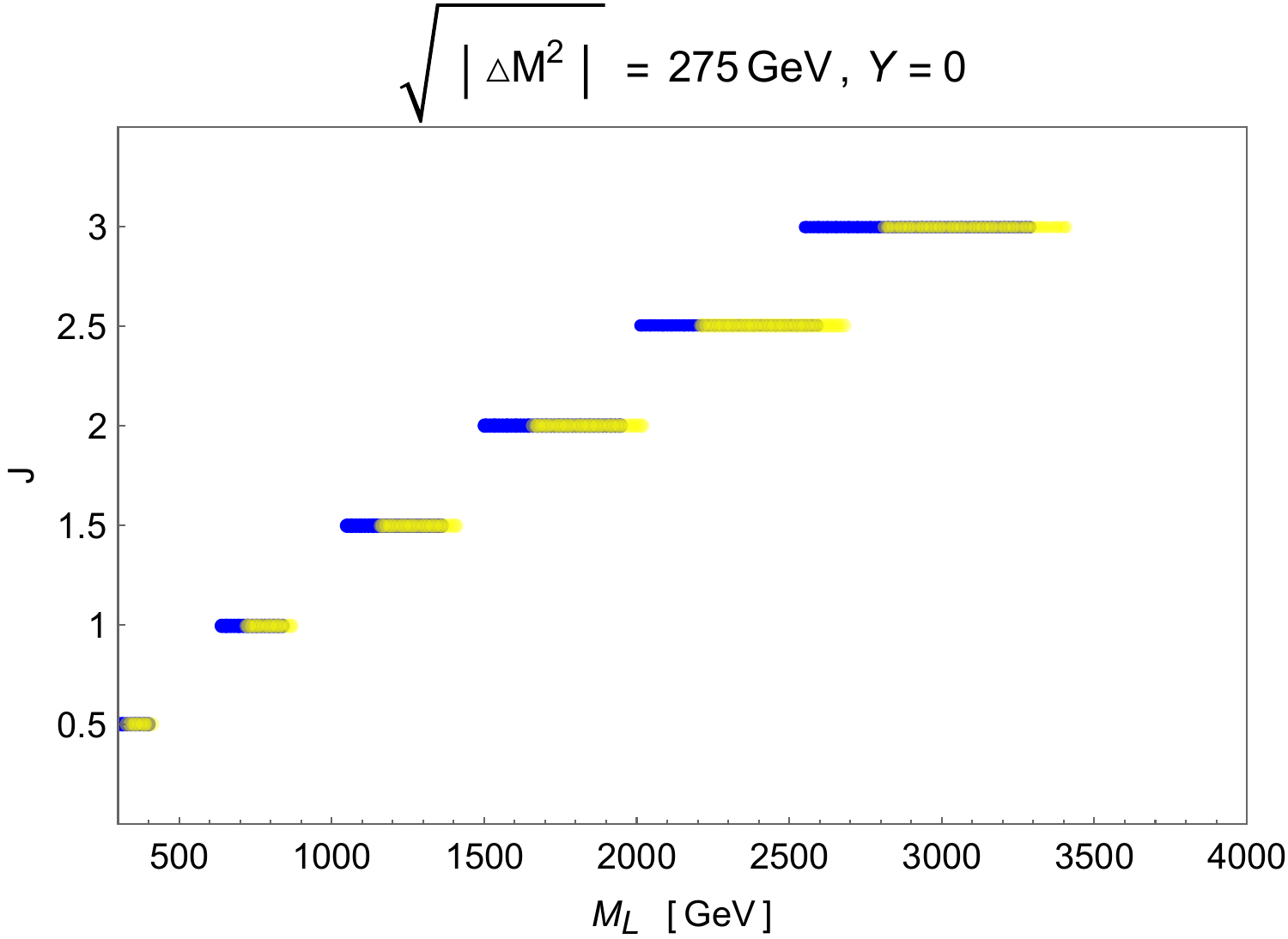}
		\includegraphics[width=0.48\linewidth]{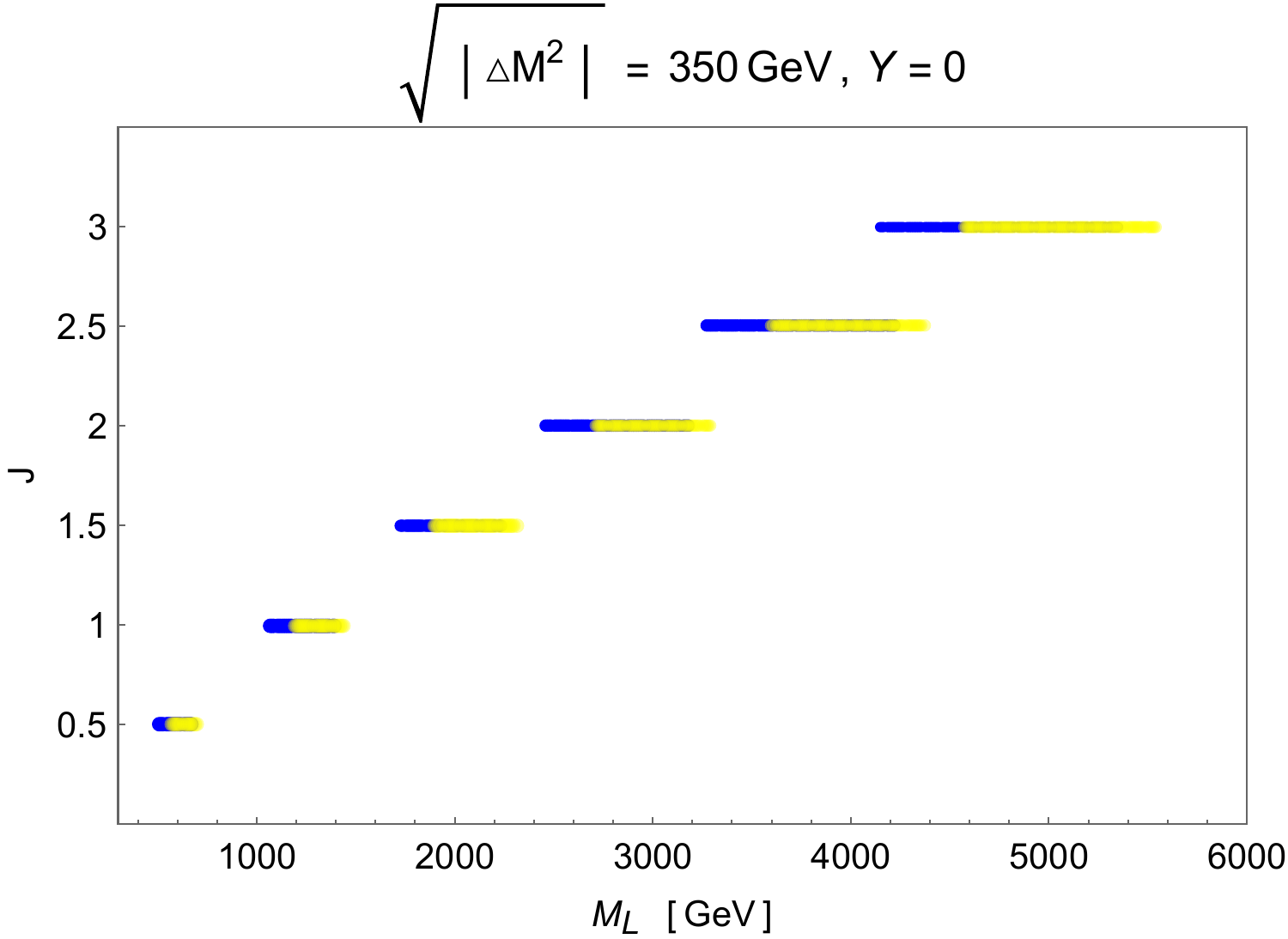}
	\end{center}
	\caption{The parameter spaces in the $M_{L}$-$J$ plane for models with scalar multiplets of $Y=0$ and $\sqrt{|\Delta M^{2}|}=$125, 200, 275 and 350~GeV  for $\lambda_{4}=1$, 2.6, 5 and 8, respectively. The color codings are the same as in Fig.~\ref{figY01}, where all the plots satisfy the pertubative limit as shown by the values of $\lambda_4$. 
	}\label{figY02}
\end{figure}

Fig.~\ref{figY01} illustrates the parameter spaces in the cases with $Y=0$ and $J=$ 1/2, 1, 2, 3. The horizontal axis labeled as $M_{L}$ represents the mass of the lightest particle in the multiplet {in the range from 300 GeV to 3000 GeV}, and the vertical axis denotes the mass difference between adjacent components. In each plot, the yellow shaded area shows the parameter space allowed by the EW global fit for $T$ in the $2 \sigma$ C.L. range when $S=U=0$, which is obtained from Ref.~\cite{Cheng:2022hbo}, while the solid blue region accords with the $2\sigma$ range of the $W$-boson mass measured by ${\rm CDF\mbox{-}\uppercase\expandafter{\romannumeral2}}$. 
The red line represents perturbative upper limit of $| \lambda_{4} | \lesssim 10 $, which can be directly related to the mass splittings of the scalar multiplet according to Eq.~(\ref{DMY0}). 
As a result, it is seen that there is always an overlap of the yellow and blue regions, meaning that most parameters explaining the $M_W$ anomaly would be consistent with the EW precision tests in the models with a scalar multiplet of $Y=0$, no matter what values of $J$. However, the required mass difference inside the mutiplet reduces with the increase of $J$, so that the perturbative limit becomes looser. Especially, the perturbativity imposes strong constraints on the lightest scalar mass to be smaller than $\sim 900~(1800)$~GeV when the weak isospin is chosen to be $J=1/2~(1)$. We can also plot the parameter spaces in the $M_L$-$J$ plane 
as shown in Fig.~\ref{figY02}. 
It is found that, with the same mass difference, the lightest mass $M_L$ in the scalar multiplet is positively correlated with its value of $J$. Moreover, with the increase of the mass splitting $\sqrt{\Delta M^2}$, the scalar mass tends to be larger with a fixed $J$. In sum, according to  Figs.~\ref{figY01} and \ref{figY02}, there is sufficient parameter spaces for a model with an extra complex scalar multiplet of $Y=0$ to explain the $W$-boson mass excess without disturbing the relevant experimental and theoretical bounds. 


\subsection{Complex Multiplet with $Y=J$}
\label{sec3.3}
In this subsection, we pay our attention to the models in which a complex scalar multiplet with $Y=J$ ($J \geq 1/2$) is included in the particle spectrum. In such a kind of models, the mass differences among scalars arise also from the operator $O_4$ as in Eq.~(\ref{DMY0}). Depending on the sign of $\lambda_4$, the model can be divided into two types:
\begin{itemize}
\item[Type] A: if $\lambda_4 >0$, the lightest particle is the most charged one in the multiplet with $M_L = M_C$, where $M_L$ denotes the mass of the lightest scalar in $\Phi_{JY}$ while $M_C$ the mass of the most electrically charged particle. By taking the case with $J=Y=2$ for example, the mass ordering is $M_{\Phi^0} > M_{\Phi^+} > M_{\Phi^{++}}> M_{\Phi^{+++}}> M_{\Phi^{++++}} \equiv M_C$, where we just denote the corresponding components by their electric charges.
\item[Type] B: if $\lambda_4<0$, the lightest scalar is the electrically neutral one in the multiplet with $M_L = M_0$, where $M_0$ stands for the mass of the neutral component. In this case, by setting $J=Y=2$, the component masses are ordered as $M_{\Phi^{++++}}> M_{\Phi^{+++}} > M_{\Phi^{++}} > M_{\Phi^+} > M_{\Phi^0} \equiv M_0$. 
\end{itemize}
With the mass splittings among the scalars in the multiplet, it is shown from Eq.~(\ref{wmass}) that the model can potentially explain the $W$-mass anomaly with nonzero corrections to the parameters $T$ and $S$ while fixing $U=0$. 

\begin{figure}[ht]
	\begin{center}
		\includegraphics[width=0.48\linewidth]{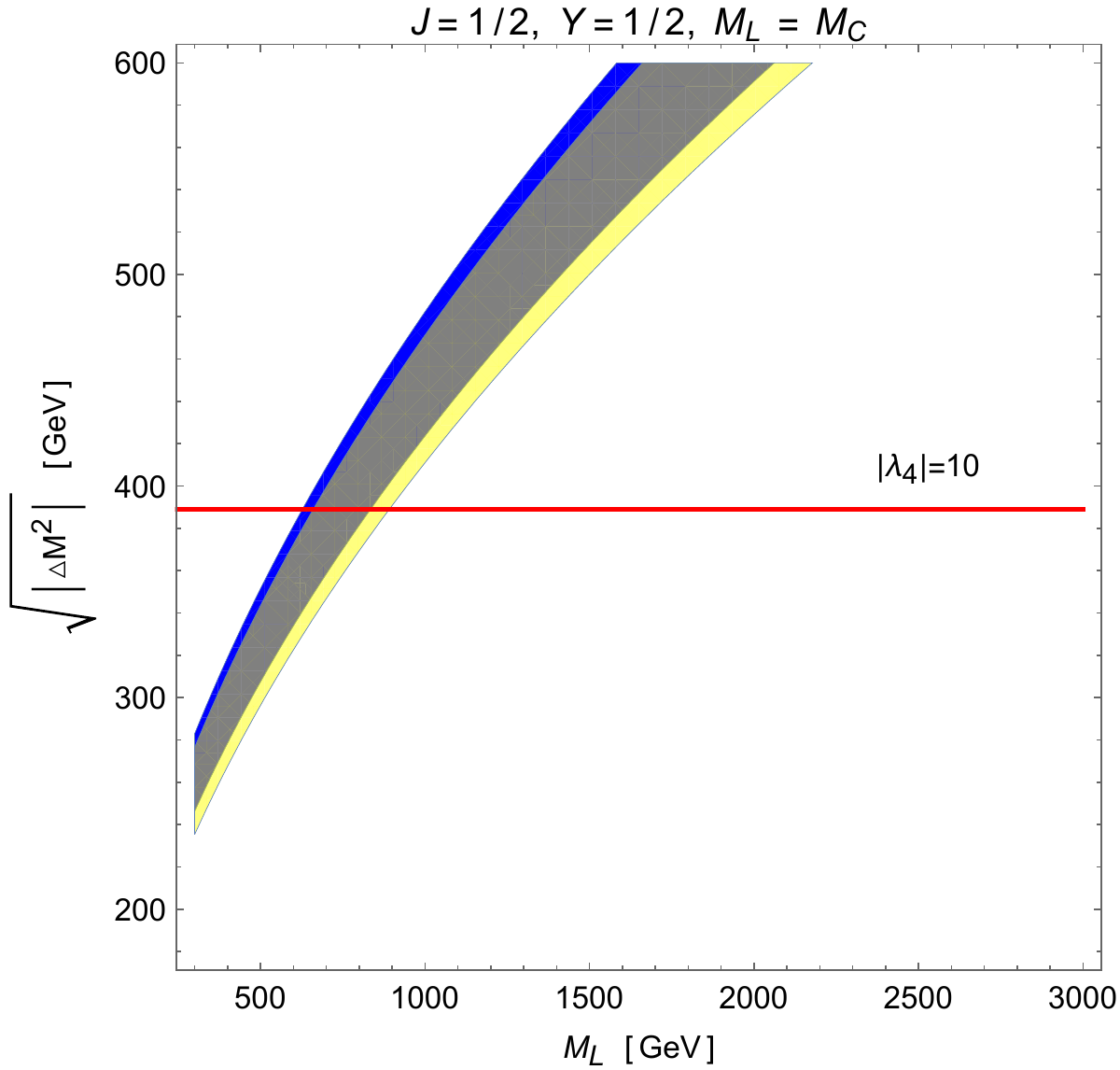}
		\includegraphics[width=0.48\linewidth]{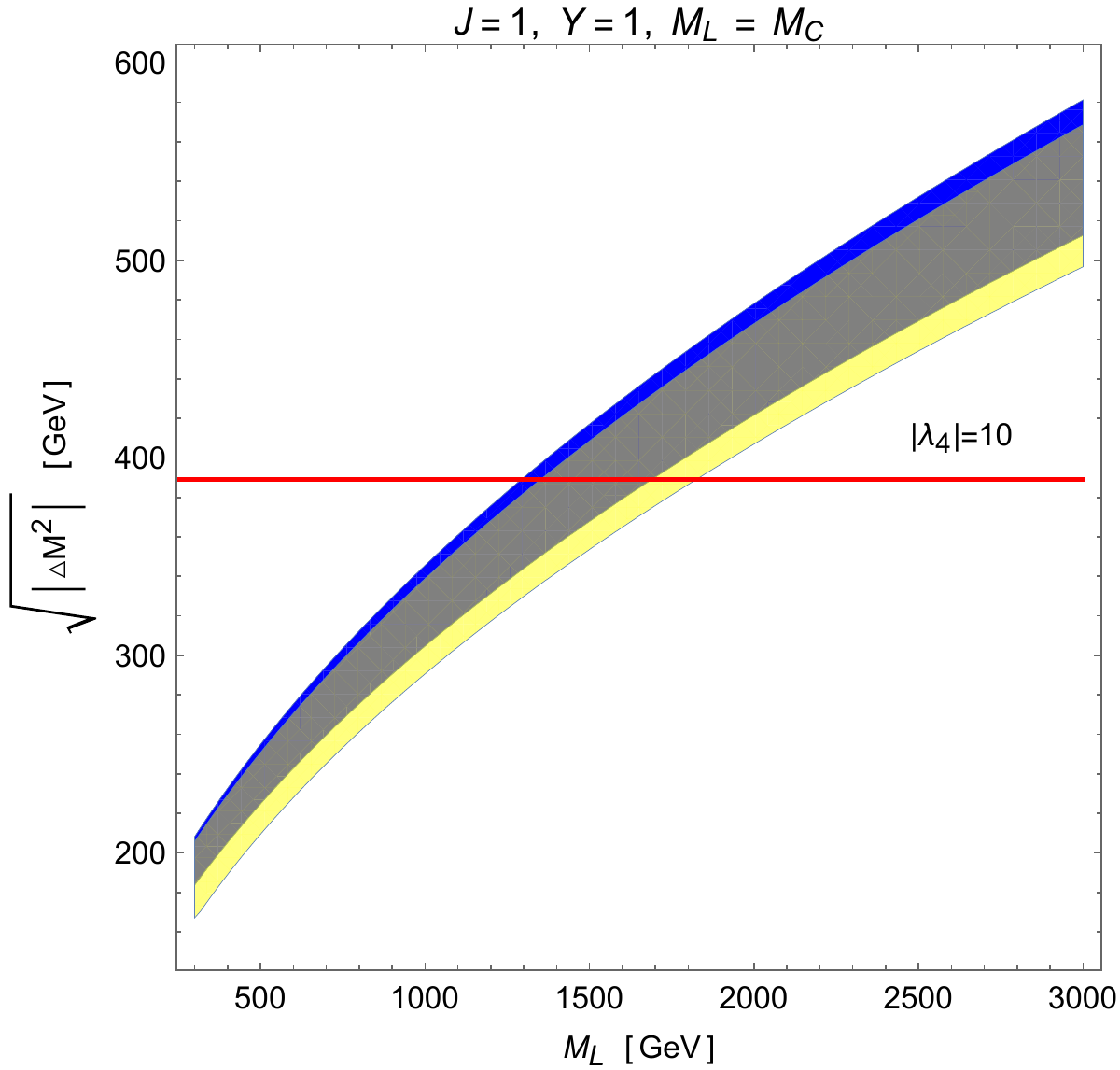}
		\includegraphics[width=0.48\linewidth]{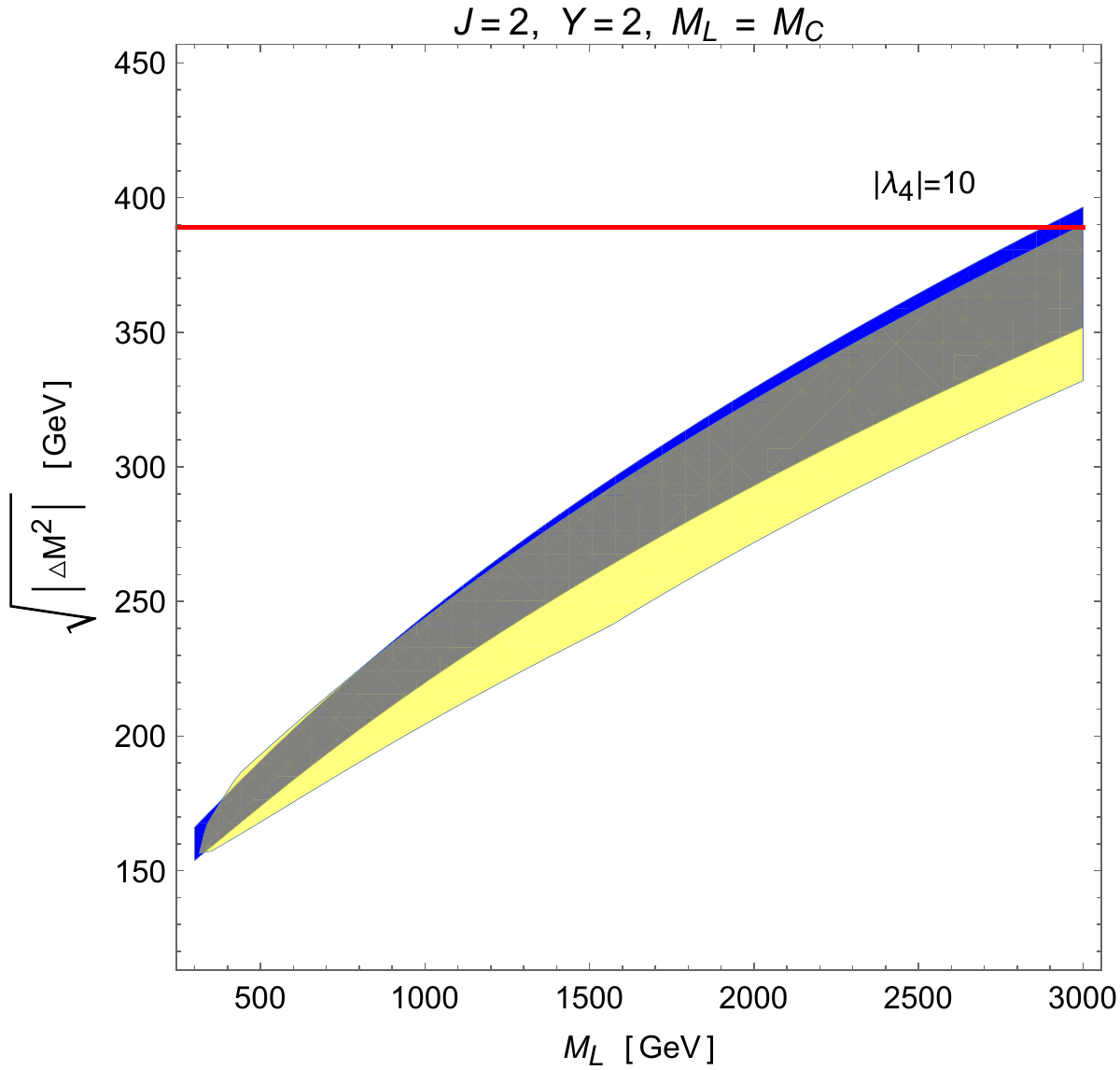}
		\includegraphics[width=0.48\linewidth]{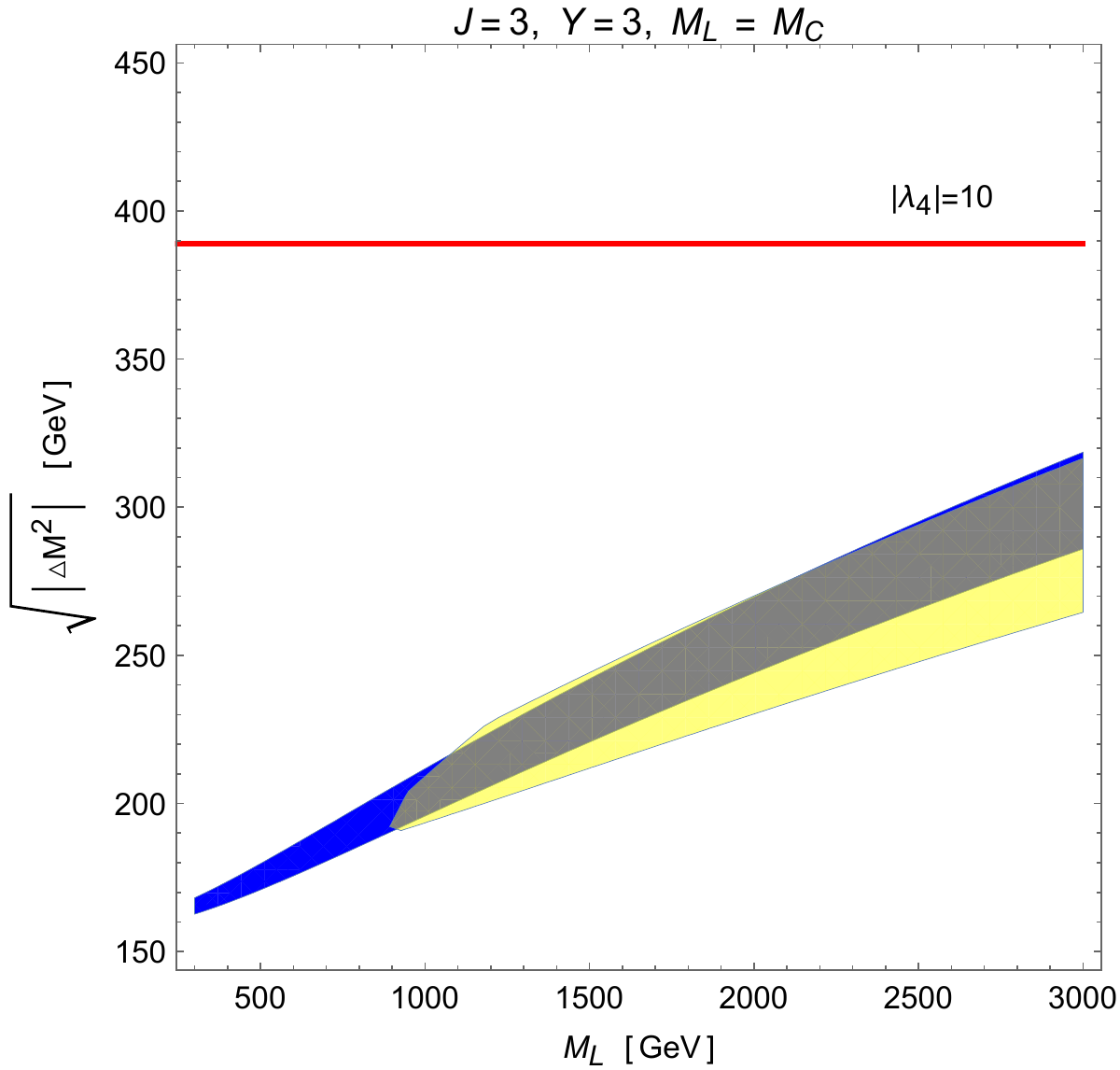}
	\end{center}
	\caption{The parameter spaces in the $M_{L}-\sqrt{|\Delta M^{2}|}$ plane for Type-A models with scalar multiplets of $Y=J=$ 1/2, 1, 2, 3, respectively. In this kind of models, the lightest scalar is the most charged one, {i.e.}, $M_{L}=M_{C}$. The color codings are the same as in Fig.~\ref{figY01}, except that the yellow area in each plot is obtained by the EW global fits of $T$ and $S$ as illustrated as the red ellipsis in Fig.~1 of Ref.~\cite{Asadi:2022xiy} .
	}\label{typeA}
\end{figure}
\begin{figure}[h]
	\begin{center}
		\includegraphics[width=0.48\linewidth]{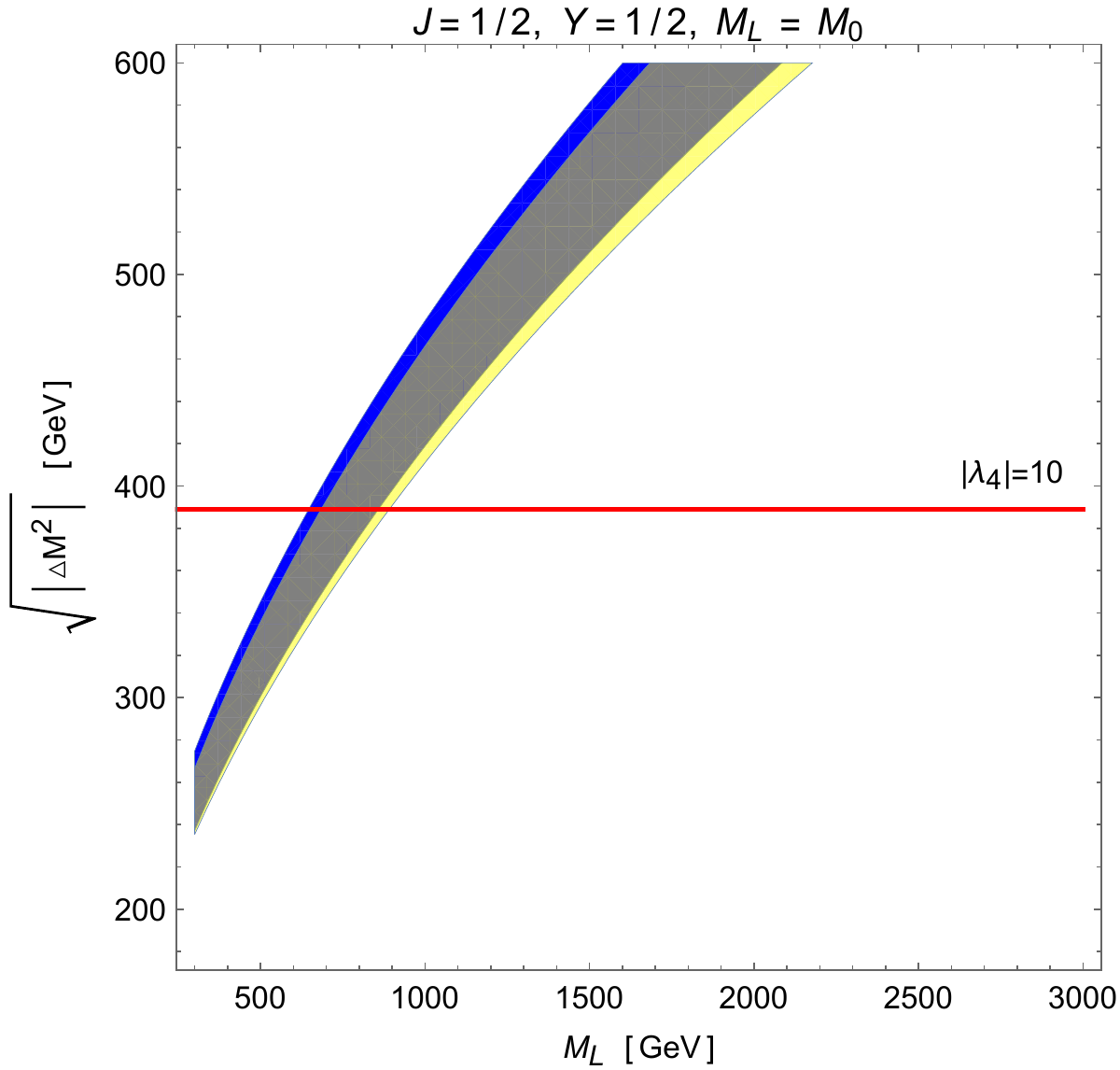}
		\includegraphics[width=0.48\linewidth]{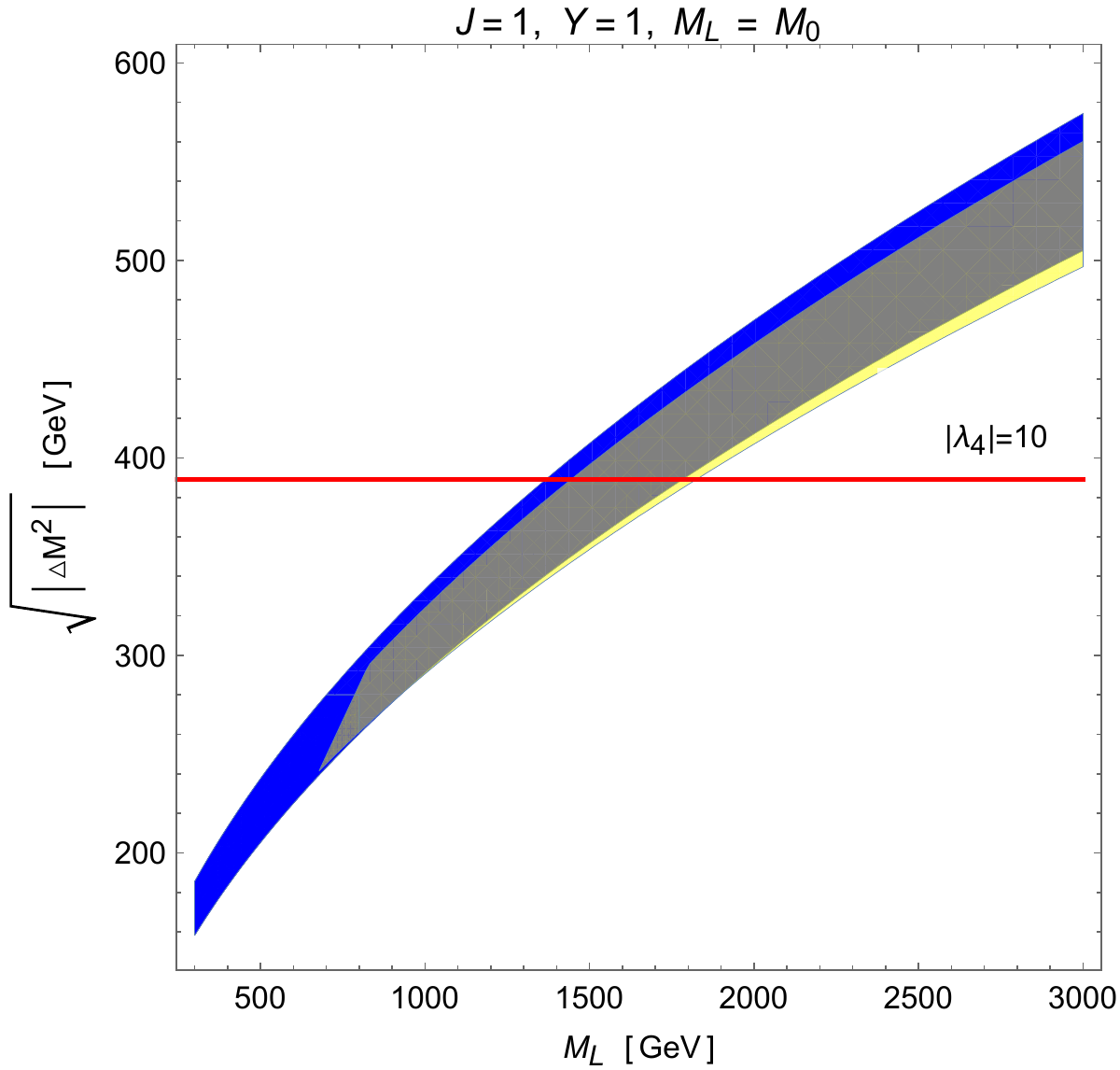}
		\includegraphics[width=0.48\linewidth]{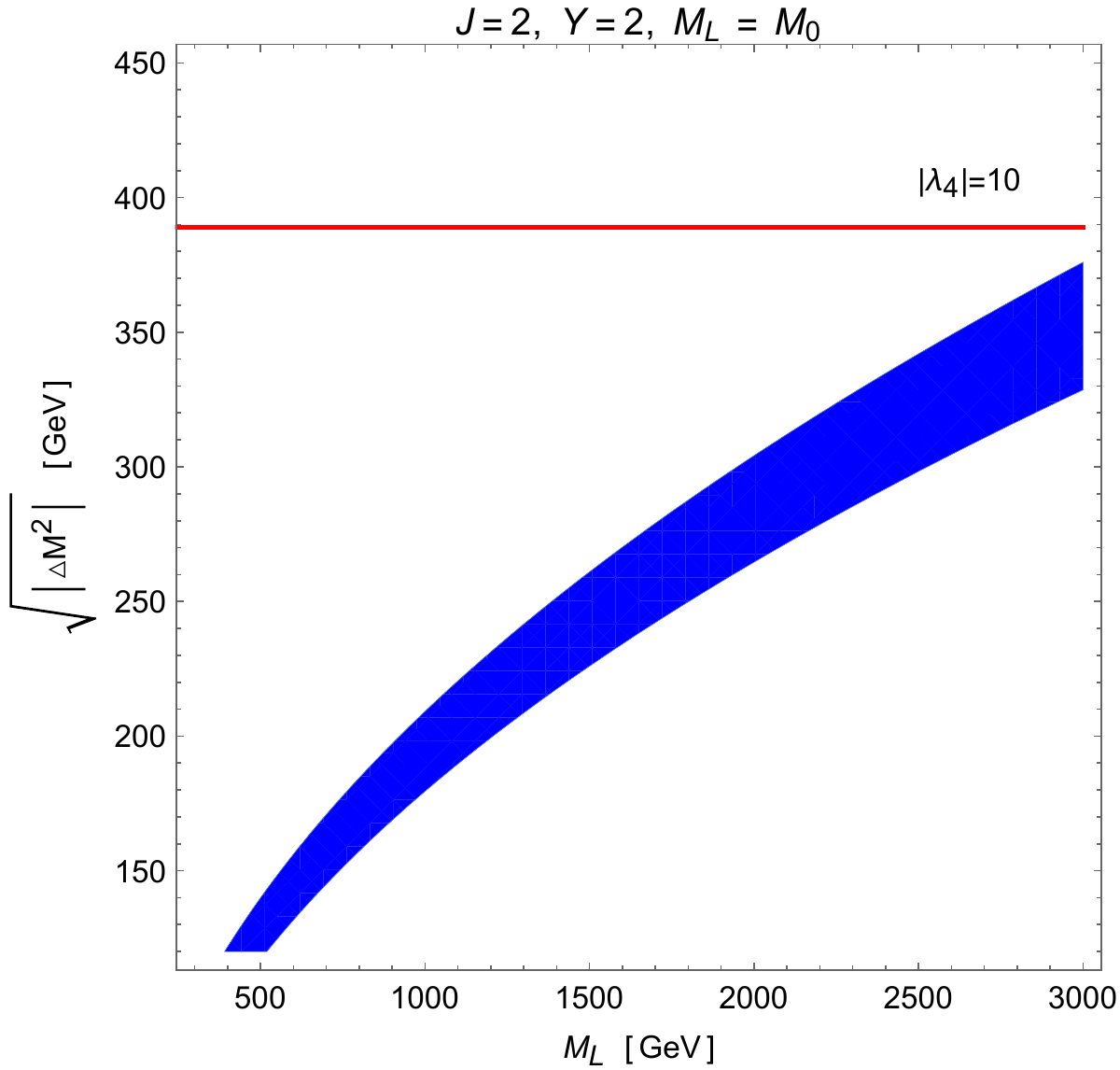}
		\includegraphics[width=0.48\linewidth]{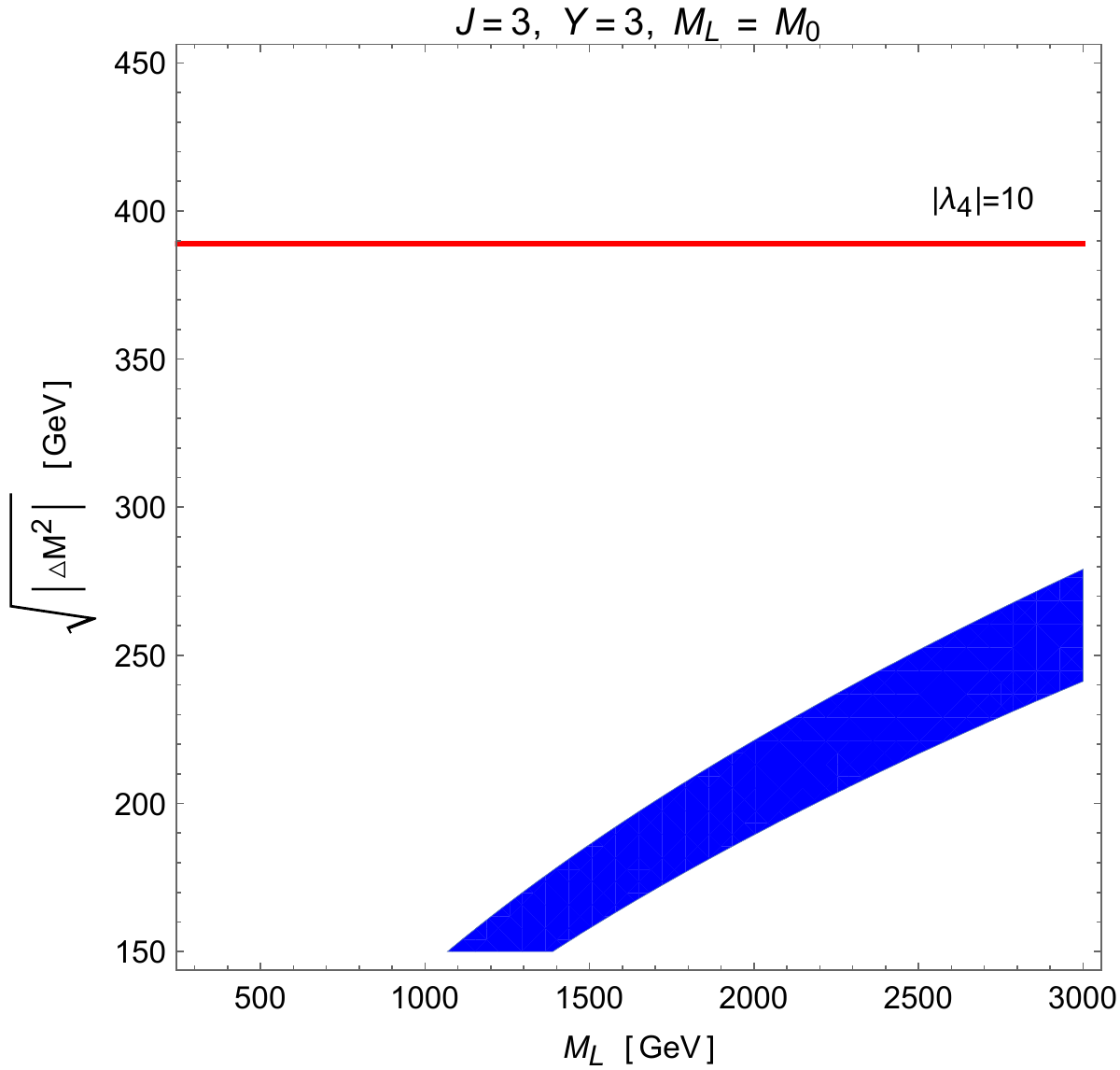}
	\end{center}
	\caption{Legend is the same as Fig.~\ref{typeA} but for Type-B models.
}\label{typeb}
\end{figure}

Figs.~\ref{typeA} and \ref{typeb} show the parameter spaces in the $M_W$-$\sqrt{\Delta m^2}$ plane for the Type-A and Type-B models by setting $Y=J= 1/2$, 1, 2, and 3, respectively. The color codings are the same as those in Fig.~\ref{figY01}, except that here the $2\sigma$ CL constraints on the oblique parameters $T$ and $S$ are obtained by the EW global fits illustrated as the red ellipsis in Fig.~1 of Ref.~\cite{Asadi:2022xiy}. It turns out that, for all Type-A models of physical interest, there are always ample available parameter spaces to solve the CDF-II $M_W$ anomaly while allowed by the EW global fits and perturbativity. Moreover, for $J=Y= 1/2~(1)$, the validity of the perturbative calculations limit the mass of the lightest scalar component as $M_L \lesssim 900~(\lesssim 1500)$~GeV, while, when $J=Y\geq 2$, the perturbativity does not provide us any useful constraint at all due to the decrease of the required mass differences. On the other hand, as seen in Fig.~\ref{typeb}, the situation changes greatly for Type-B models. In this case, the CDF-II regions to explain the $M_W$ excess are totally excluded by the global fits of various EW precision observables for all models with $J=Y\geq 2$. For the scalar triplet with $J=Y=1$, compared with its Type-A counterpart, the low-mass region with $M_L\lesssim 700$~GeV and the high-mass region with $M_L \gtrsim 1800$~GeV are disfavored by the EW global fits and the perturbativity, respectively. Finally, constrained by the perturbativity upper bound on $\lambda_4$, the model with a EW doublet of $J=Y=1/2$ now admits the parameter space with $M_L \lesssim 900$~GeV to account for the $M_W$ signal observed by CDF-II.  


\subsection{Effects of Varying $Y$}
\label{sec3.4}
\begin{figure}[ht]
	\begin{center}
		\includegraphics[width=0.48\linewidth]{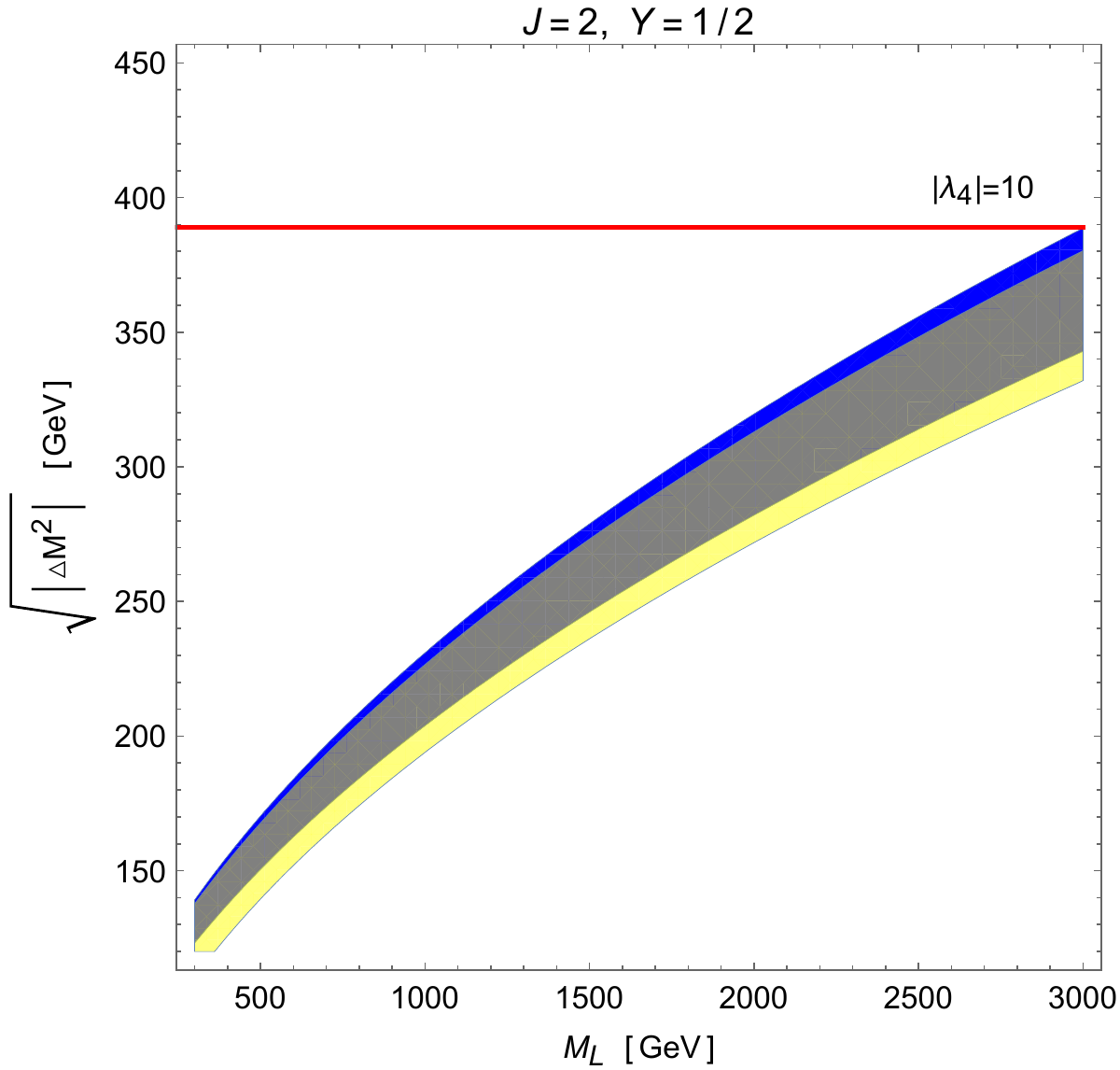}
		\includegraphics[width=0.48\linewidth]{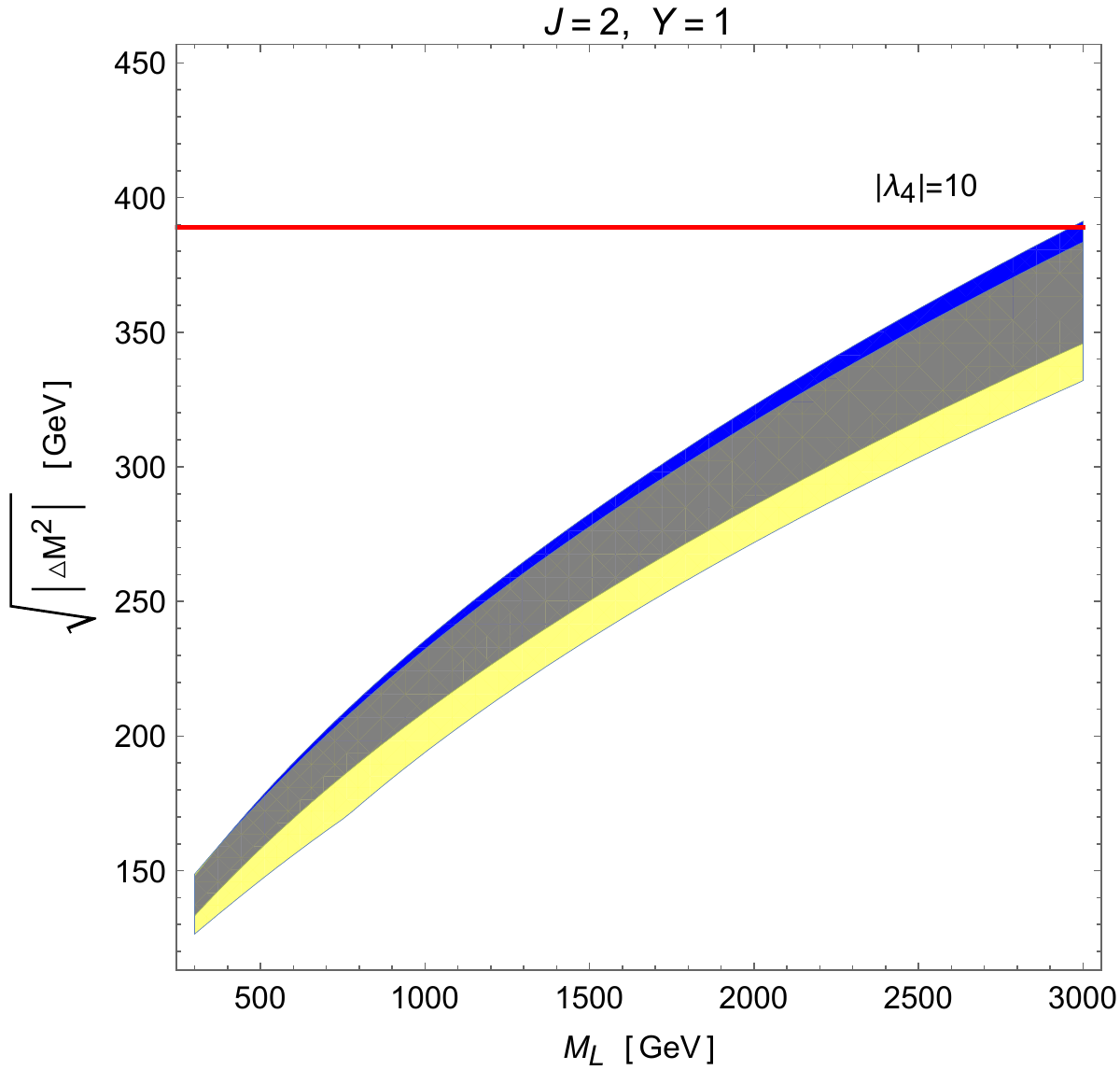}
		\includegraphics[width=0.48\linewidth]{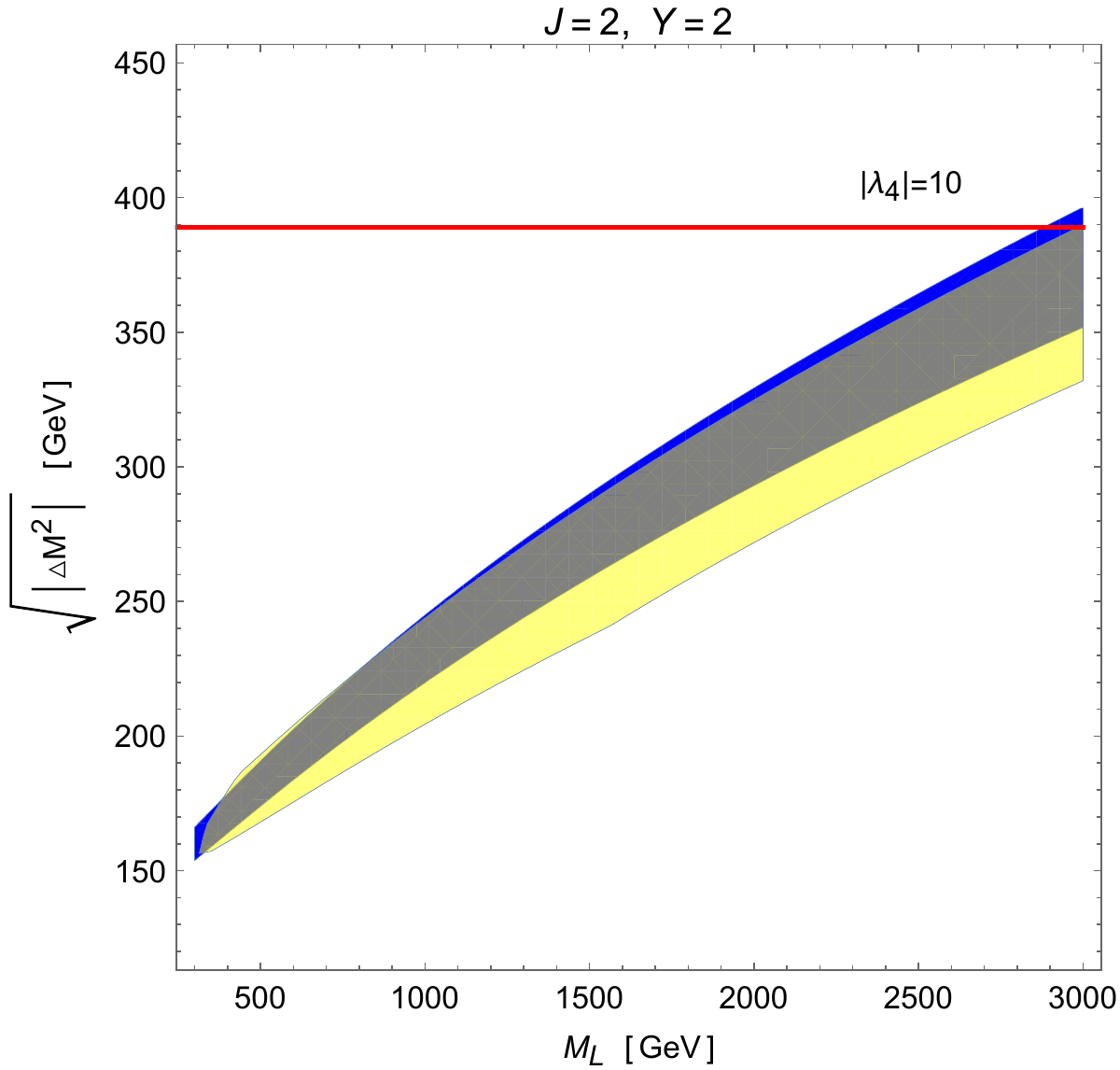}
		\includegraphics[width=0.48\linewidth]{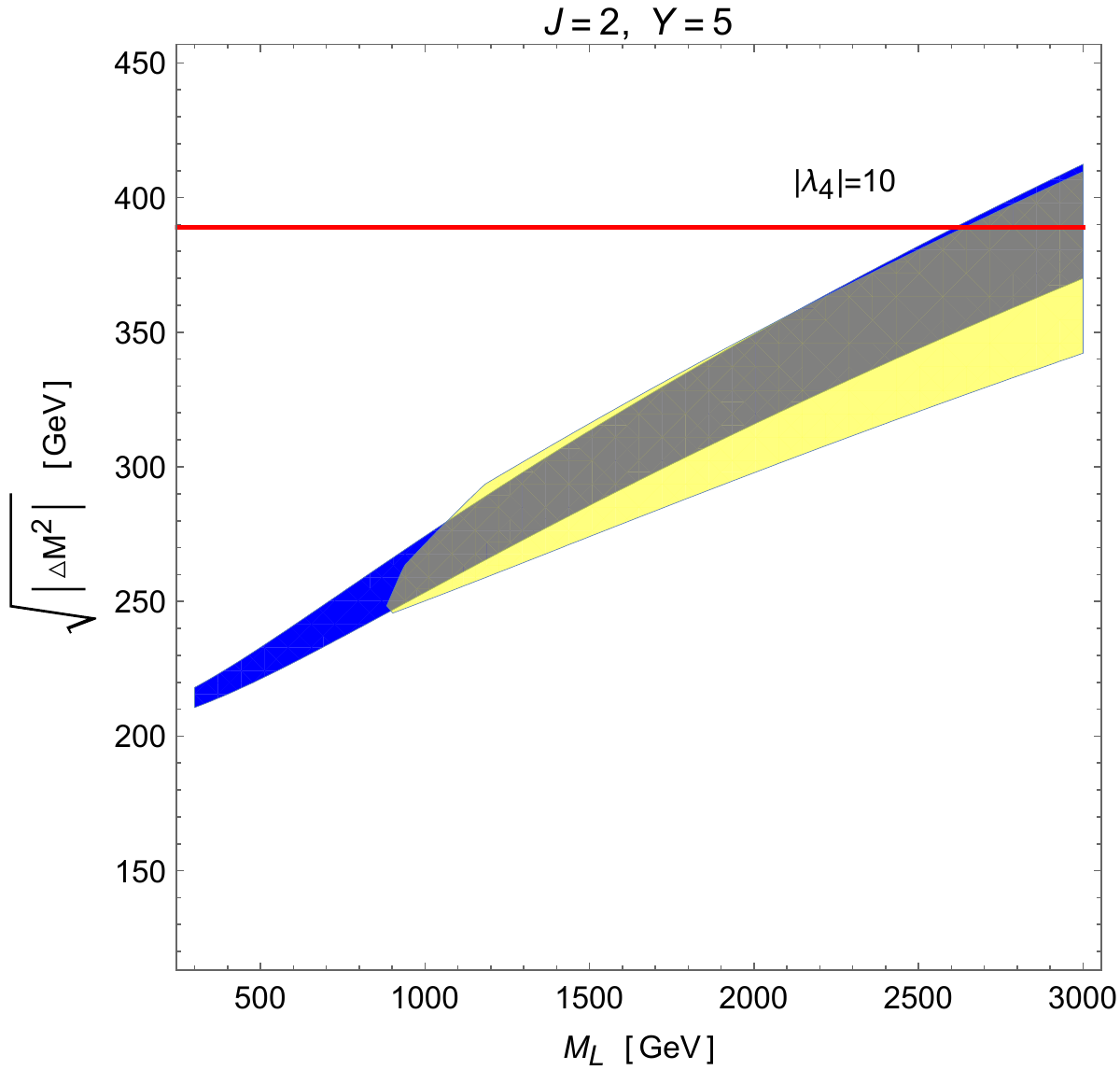}
	\end{center}
	\caption{Legend is the same as Fig.~\ref{typeA} but for
	models with an additional scalar multiplet of $J=2$ and $Y=$ $1/2$, 1, 2, and 5, respectively. 
	}\label{figYPJ2}
\end{figure}

\begin{figure}[h]
	\begin{center}
		\includegraphics[width=0.48\linewidth]{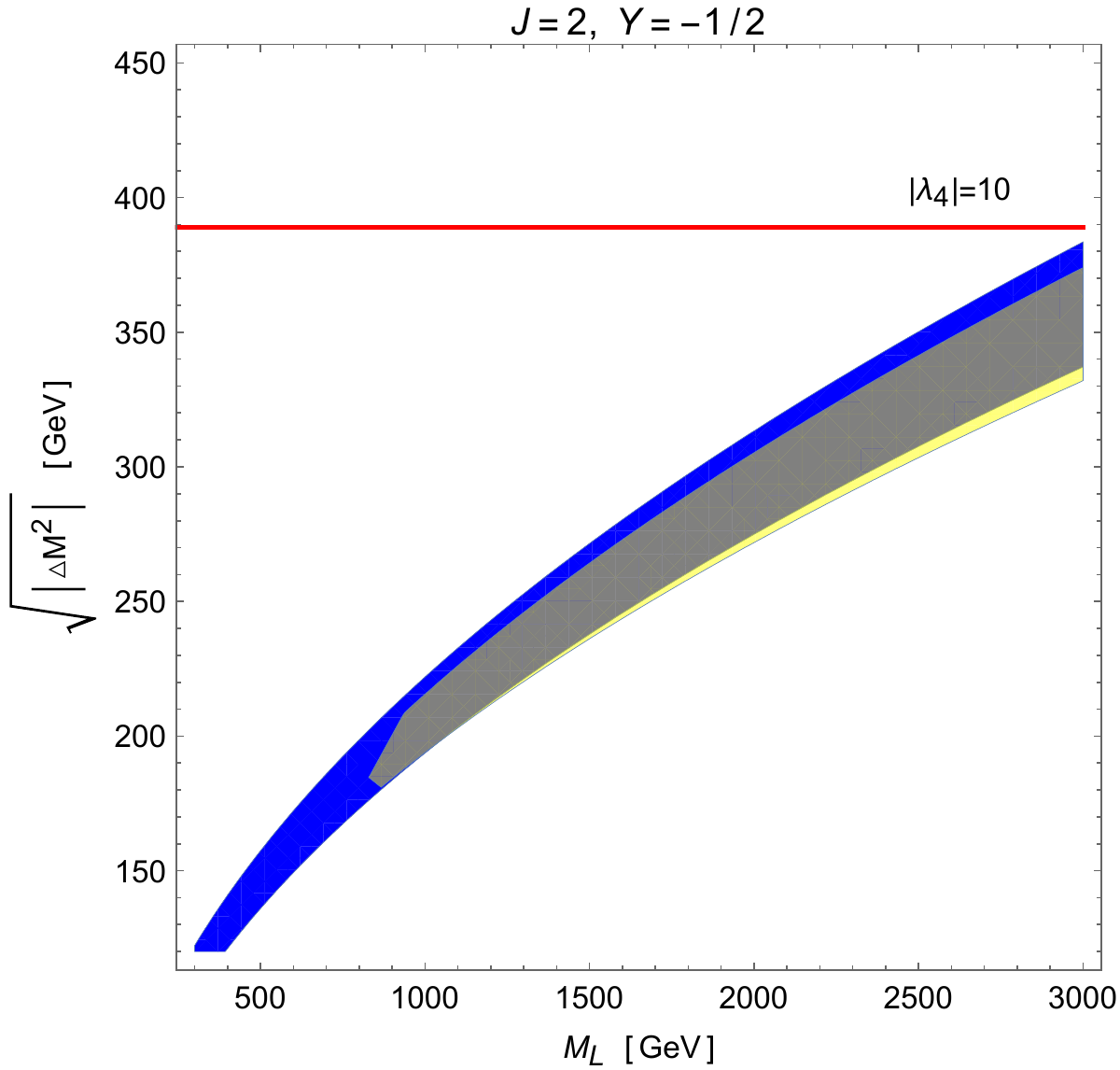}
		\includegraphics[width=0.48\linewidth]{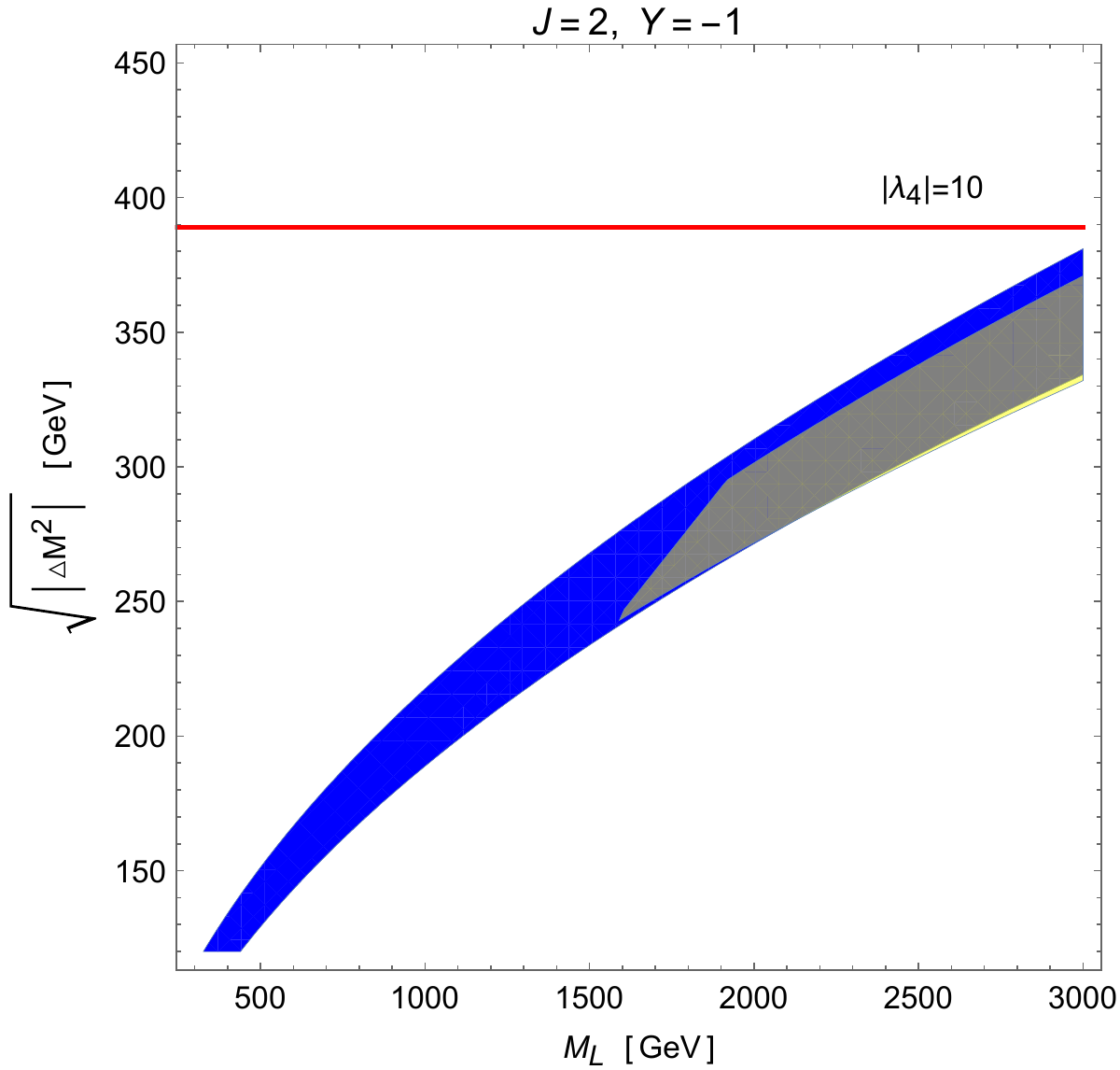}
		\includegraphics[width=0.48\linewidth]{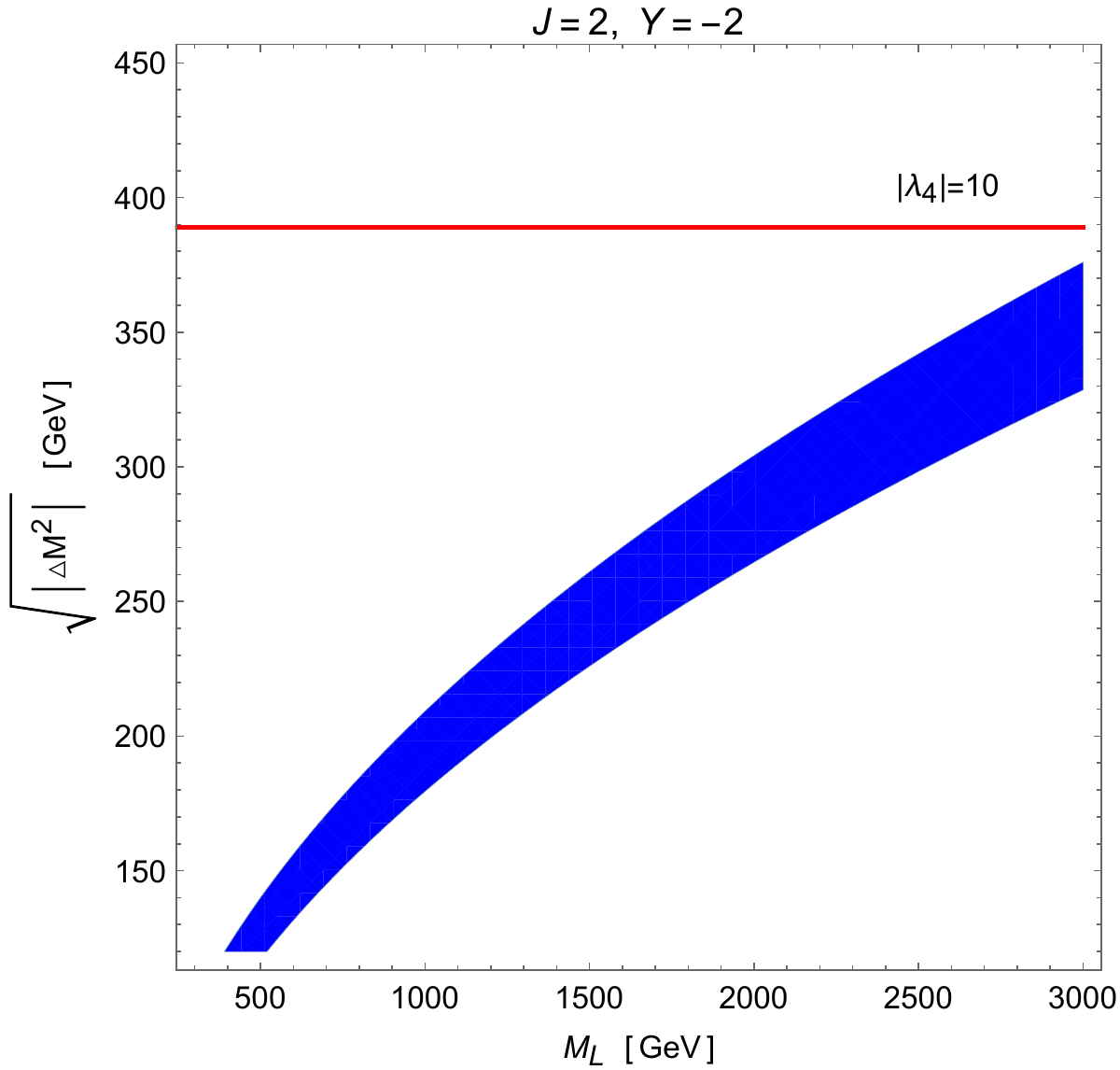}
		\includegraphics[width=0.48\linewidth]{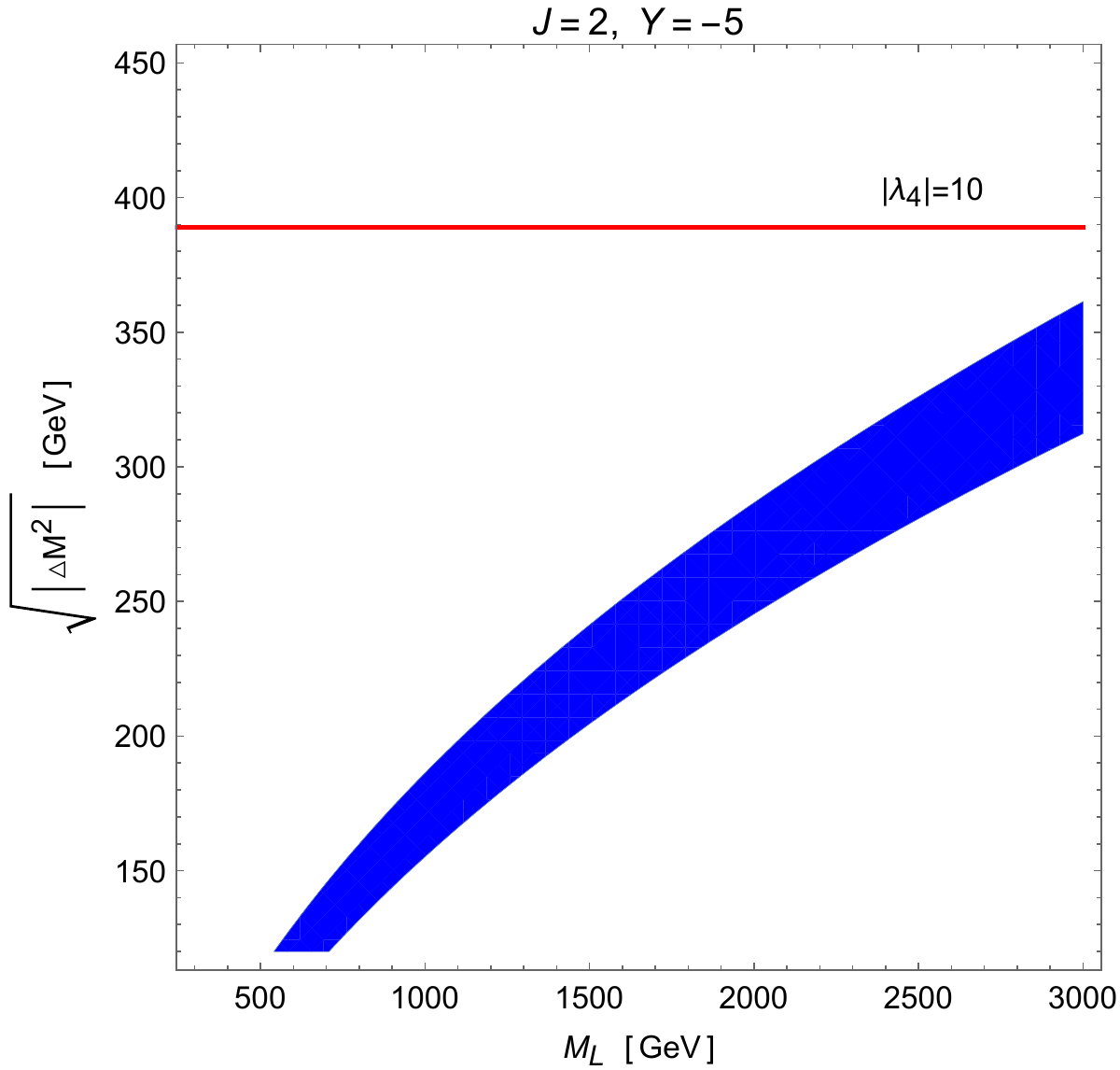}
	\end{center}
	\caption{Legend is the same as Fig.~\ref{typeA} but for
	for models with an additional scalar multiplet of $J=2$ and $Y=$ $-1/2$, -1, -2, -5, respectively. 
	}\label{figYMJ2}
\end{figure}

In the light of Secs.~\ref{sec3.2} and \ref{sec3.3}, we find that the value of $Y$ might have a significant influence on the scalar multiplet explanation to the CDF-II $M_W$ anomaly. For the sake of comprehensiveness, we would like to investigate the effects of the variation of $Y$ on the allowed parameter space in this subsection. To be concrete, we consider the model including an extra scalar multiplet with $J=2$, $\lambda_4>0$ and $Y=\pm 1/2$, $\pm 1$, $\pm 2$, and $\pm 5$, respectively. As shown below, with $\lambda_4 >0$, the phenomenology of models with positive $Y$ is quite different from that with negative $Y$, so that we show these two cases separately as in Figs.~\ref{figYPJ2} and \ref{figYMJ2}, respectively. The color codings in these two plots are the same as in Fig.~\ref{figY01}.



According to Fig.~\ref{figYPJ2}, we find that the mass splitting between two neighbouring components required to solve the $W$-mass anomaly increases gradually with $Y$ becoming larger in all range of $M_L$. Also, in all cases, most parameter spaces explaining the $M_W$ excess can be accommodated by the EW global fits, except the low-mass regions with $M_L \lesssim 300~(800)$~GeV for the benchmarks with $Y=2~(5)$.   Furthermore, the upper limit from the perturbativity of $\lambda_4$ only gives a relevant constraint on the parameter regions with large values of $M_L$. In particular, the perturbative limit rules out the portion with $M_L \gtrsim 2900~(2500)$~GeV for the multiplet with $Y=2~(5)$. 

In contrast, as shown in Fig.~\ref{figYMJ2},
the parameter spaces allowed by all experimental data are greatly reduced when $Y$ becomes more negative, especially for the low $M_L$ regions. What is more, all the blue bands in the cases $Y \leqslant -2$ which  explains  the CDF-II measured values of $M_W$ are now totally excluded by the EW precision data. From our working experiences, the experimental limit of the oblique parameter $S$ gives the dominant constraint to the models with negative $Y$'s. Actually, we can understand this result by inspecting the expression of $S$ in the limit of small mass splittings from $O_4$ as follows:
\begin{eqnarray}\label{SApp}
	S &=&-\frac{Y}{3\pi}\left(2\ln M_{\varPhi_{2}}^{2}+\ln M_{\varPhi_{1}}^{2}-\ln M_{\varPhi_{-1}}^{2}-2\ln M_{\varPhi_{-2}}^{2}\right) \nonumber\\
	&=&-\frac{Y}{3\pi}\left[2\ln\left(\frac{M_{\varPhi_{-2}}^{2}-\lambda_{4}v^{2}}{M_{\varPhi_{-2}}^{2}}\right)+\ln\left(\frac{M_{\varPhi_{-1}}^{2}-\lambda_{4}v^2/2 }{M_{\varPhi_{-1}}^{2}}\right)\right] \nonumber\\
	&\approx & \frac{\lambda_{4}Y}{3\pi}\left(\frac{2v^{2}}{ M_{\varPhi_{-2}}^{2}}+\frac{v^{2}}{2 M_{\varPhi_{-1}}^{2}}\right)\,,
\end{eqnarray}
where the subscript $I$ of the scalar $\Phi_I$ denotes the third isospin value of the component whose electric charge can be obtained by $Q=I+Y$. The last relation is the approximation when $ |\lambda_4 v^2| \ll M_{\Phi_I} $ for any $I$. It is seen from Eq.~(\ref{SApp}) that, for $\lambda_4 >0$, the negative value of $Y$ would lead to $S < 0$. According to the EW global fit shown as the red circle in Fig.~1 in Ref.~\cite{Asadi:2022xiy}, the current EW precision data together with the new CDF-II measurement of $M_W$ has constrained $-0.04 \lesssim S \lesssim 0.36$ at 2$\sigma$ CL. This means that a multiplet scalar with $Y < 0$ would suffer from a constraint of $S$ much stronger than that with $Y>0$, which explains the differences shown in Figs.~\ref{figYPJ2} and \ref{figYMJ2}.      



\section{Conclusions and Discussions}
\label{sec4}
In the light of the recent measurement of the $W$-boson mass by the CDF-\uppercase\expandafter{\romannumeral2} Collaboration, we have comprehensively studied the explanation of this $M_W$ anomaly in terms of one-loop effects of a  general $SU(2)_{L}$ scalar multiplet.
As shown in the literature, in the case without scalar VEVs, the dominant contribution to $M_W$ can be expressed at  leading order as the linear combination of the oblique parameters $T$ and $S$, which would be constrained by the global fits of various EW precision observables. Also, it is found that the operator $O_4$ gives rise to the main contribution to the mass splittings among components in the multiplet, which is needed to generate nonzero corrections to $T$ and $S$. Thus, its coefficient $\lambda_4$ would be limited by the perturbativity. In the present work, we have rederived the general formulae for the one-loop contributions to $T$ and $S$ from a scalar multiplet, confirming the results in the literature. We have applied these analytic expressions to several models in order to explore the effects of the multiplet isospin representations and hypercharges on the viability to explain the CDF-\uppercase\expandafter{\romannumeral 2} $M_W$ anomaly. As a result, for a scalar under the $SU(2)_L$ real representation with $Y=0$, the model cannot explain the $M_W$ excess {at the one-loop level} due to the vanishing of the mass splitting between adjacent scalar components. In contrast, the mass differences would be induced by $O_4$ for a general complex representation, so that such models would potentially solve the $M_W$ discrepancy between SM calculations and CDF-\uppercase\expandafter{\romannumeral 2} values. 
Concretely, for the mutliplets with $Y=0$, the $M_W$ excess can be explained solely by the corrections of $T$ due to $S = 0$. All the models with $J=1/2$, 1, 2, and 3 are shown to have sufficient parameter spaces to meet the ${\rm CDF\mbox{-}\uppercase\expandafter{\romannumeral2}}$ result and the EW global fits. 
For the cases with $Y=J$, the phenomenology can be divided into two classes depending on the sign of $\lambda_4$. If $\lambda_4 > 0$, the models are labeled as Type-A in which the lightest scalar is the most charged one. It turns out that there are always parameter spaces that can accommodate the CDF-\uppercase\expandafter{\romannumeral 2} $M_W$ value while still agreeing with the EW global fits and the perturbativity limit. On the other hand, for the Type-B models with $\lambda_4 < 0$, the lightest particle is the electrically neutral scalar. The parameter region simultaneously allowed by the CDF-\uppercase\expandafter{\romannumeral 2} and EW precision test data shrinks greatly as the increase of $J=Y$. In particular, when $J=Y\geqslant 2$, all regions with the lightest scalar mass below 3000~GeV are ruled out by the EW precision tests. In addition, we have  investigated the effects of the hypercharge $Y$ on the scalar multiplet solution to the $M_W$ anomaly. We fix $J=2$ and take $\lambda_4 >0$. When $Y$ is positive, the parameter spaces always exist for the interpretation of the $M_W$ excess while allowed by other constraints. However, for $Y<0$, it is seen that only the cases with $ -1 \leq Y<0$ allow parameter spaces for the viable explanation to the $W$-boson mass anomaly, whereas for $Y\leqslant -2$, all CDF-\uppercase\expandafter{\romannumeral 2} favored regions are excluded by the constraint on the oblique parameter $S$. 

{Note that it was argued in Ref.~\cite{Lavoura:1993nq} that when the multiplet scalar masses are light or the multiplet representation $J$ is high, the oblique parameter $U$ would make a sizable contribution to the W-boson mass. Concretely, for the high-multiplet cases with $J=2$ and $J=3$, we find that, when the scalar particles are light enough, say the mass of the lightest particle is only$M_L=$200 GeV, $U$ will give the $W$ mass correction as large as 10\%-15\% of that of $T$ or $S$. This means that, even though $T$ and $S$ still dominate over the $M_W$ correction, the effect of $U$ cannot be ignored. However, when $M_L$ raises to above 500 GeV, the $U$ effect to $M_W$ is much suppressed, and only contributes to 1\% $M_W$ corrections as compared with the dominant $T$ parameter. This result can be understood from the EFT perspective. After all, the $T$ and $S$ parameters come from the dimension-6 operators, while $U$ from dimension-8 one. Only when we take a low cutoff scale which can be identified as the lightest scalar mass as $\Lambda \sim M_L \sim$ 200 GeV, the suppression from the energy cutoff is not significant, and we can have an observable effect from $U$. Moreover, the effect of $U$ is also sensitive to the isospin representation $J$. We have found that when $J$ is taken an extraordinary value as $J=8$ for $M_L=$300 GeV, the contribution of $U$ to $M_W$ can be comparable to or even larger than that of $T$. Nevertheless, in the present work, the mass of the multiplet scalar is large and $J\leq 3$ is relatively small,  so that the contribution of $U$ is always suppressed compared with $T$ and $S$. } 
 
\begin{figure}[h]
	\begin{center}
		\includegraphics[width=\linewidth]{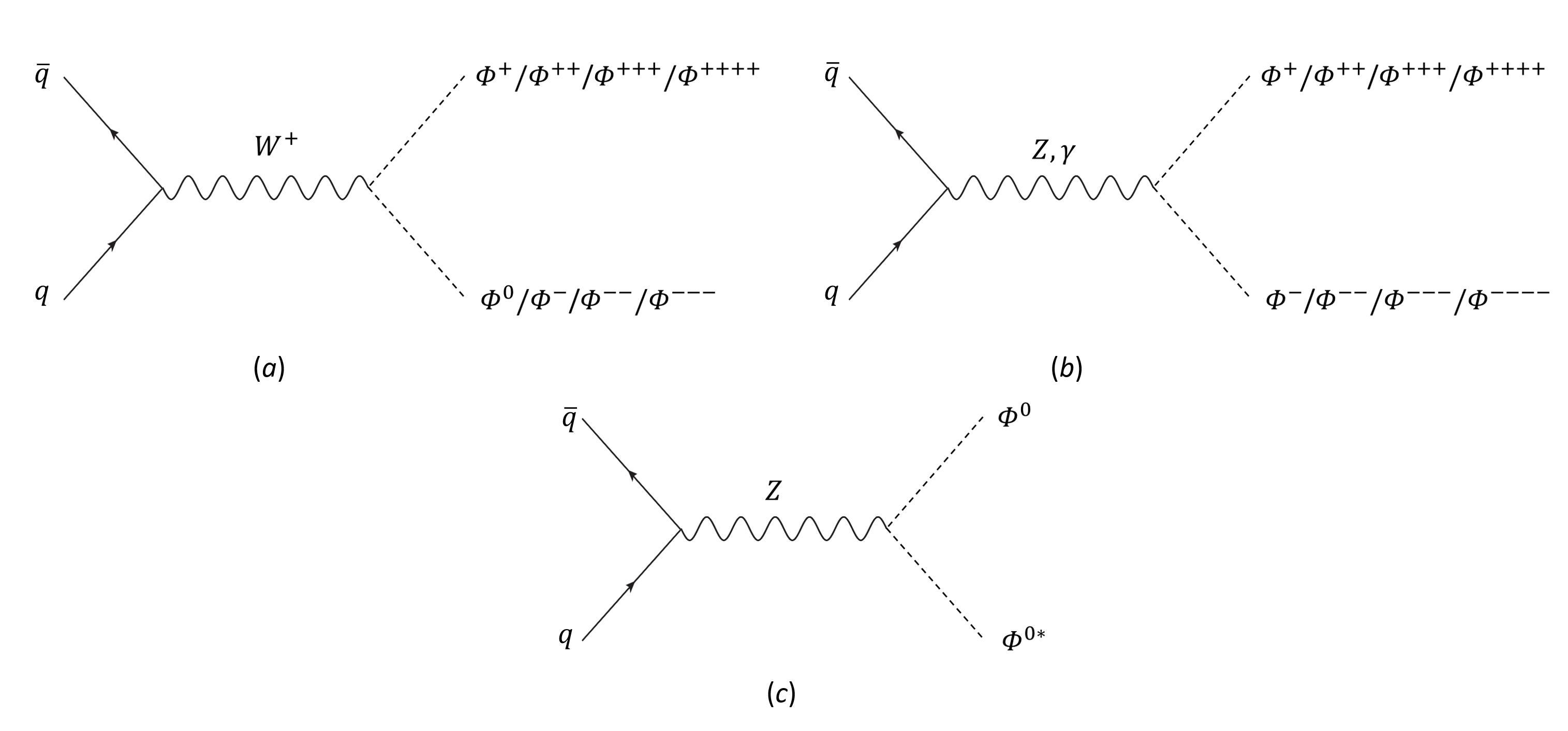}
	\end{center}
	\caption{Dominant production channels for new scalar particles in the multiplet in the Type-A models with $J=Y=2$.}\label{production}
\end{figure}

Finally, we would like to make some comments on the possible collider signatures for such scalar multiplet extensions of the SM. In particular,  note that for a high-dimensional representation, the multiplet contains highly electrically charged states in the spectrum, which would give us spectacular collider signals at the LHC. To be concrete, we take the example with an extra scalar multiplet of $J=Y=2$, in which the electric charges of scalars can be as high as $\pm 4$. Also, we focus on the case in which the mass of the lightest scalar is smaller than 1~TeV, so that it could be produced directly at the high-luminosity run of the LHC. As studied in Sec.~\ref{sec3.3}, the Type-B case in which the lightest scalar is the neutral one has been ruled out under the scrutiny of the EW precision tests. Thus, in what follows, we shall concentrate on the Type-A models in which the lightest state is the most charged one, whose electric charge is $\pm$4. In Fig.~\ref{production}, we show that the heavy scalars can be produced dominantly through the Drell-Yan processes~\cite{Drell:1969jm,Drell:1970wh,Drell:1970yt} 
via the mediation of the EW gauge bosons, such as $W$, $Z$ and photons.   
In order to study the LHC signatures of the generated multi-charged scalars, we need to specify their decay products. Note that in the Lagrangian in Eq.~(\ref{potential}), there is an accidental $Z_2$ symmetry imposed on the high-dimensional scalar multiplet $\Phi_{JY}$, which causes the lightest particle to be stable against decaying. In the Type-A multiplet with $J=Y=2$, such the lightest scalar should be $\Phi^{\pm \pm \pm \pm}$. If this state were stable, then it would become a charged DM candidate, which is not allowed by the present cosmological and astronomical observations. Therefore, we are required to break this problematic $Z_2$ symmetry. There are two ways to achieve this. Firstly, the $Z_2$ symmetry can be broken spontaneously by allowing the neutral component to possess a small VEV $v_\Phi$. Even though $v_\Phi$ would also give rise to new contributions to the $W$-boson mass at the tree level, the one-loop effects due to the modifications of $T$ and $S$ would still dominate over the $M_W$ corrections as long as $v_\Phi \lesssim 1$~GeV, which is the region of an extra EW scalar VEV allowed by the EW global fits. As a result, the heavier states in the multiplet would firstly decay into the lightest one $\Phi^{\pm \pm \pm \pm}$ via hierarchical cascades. 
\begin{figure}[ht]
	\begin{center}
		\includegraphics[width=\linewidth]{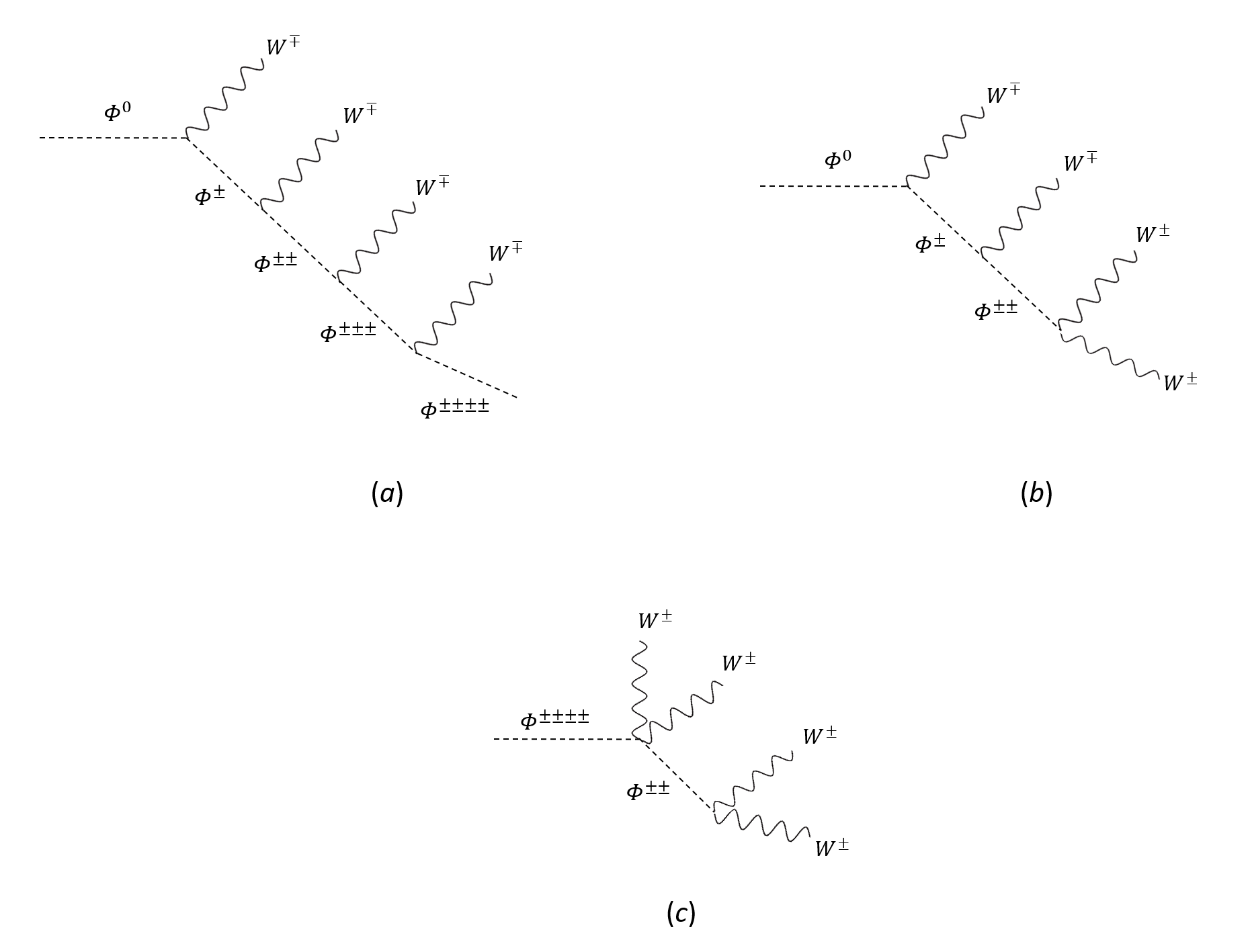}
	\end{center}
	\caption{Decay chains of the scalar particles when the scalar multiplet has a VEV in the Type-A model with $J=Y=2$.}\label{decay1}
\end{figure}
In the plot $(a)$ of Fig.~\ref{decay1}, we take for example the cascade decay chain of the heaviest scalar $\Phi^0$ by emitting multiple $W$-bosons. With the VEV $v_\Phi$, the most charged scalar which is also the lightest would further decay into 4 $W$-bosons as illustrated in the plots $(b)$ and $(c)$ in Fig.~\ref{decay1}. The other way to violate the $Z_2$ symmetry is to add some high-dimensional operators, such as the dimension-5 operator $\Phi_{abcd}(H^{*})^{a}(H^{*})^{b}(H^{*})^{c}(H^{*})^{d}$ or the dimension-7 one $\Phi_{abcd}\left(\bar{L}^C\right)^{a}\left(\hat{L}\right)^{b}\left(\bar{L}^C\right)^{c}\left(\hat{L}\right)^{d}$, where the capital letter $C$ denotes the charge conjugation while the lowercase Latin letters represent the $SU(2)_L$ fundamental indices. 
\begin{figure}[h]
	\begin{center}
		\includegraphics[width=\linewidth]{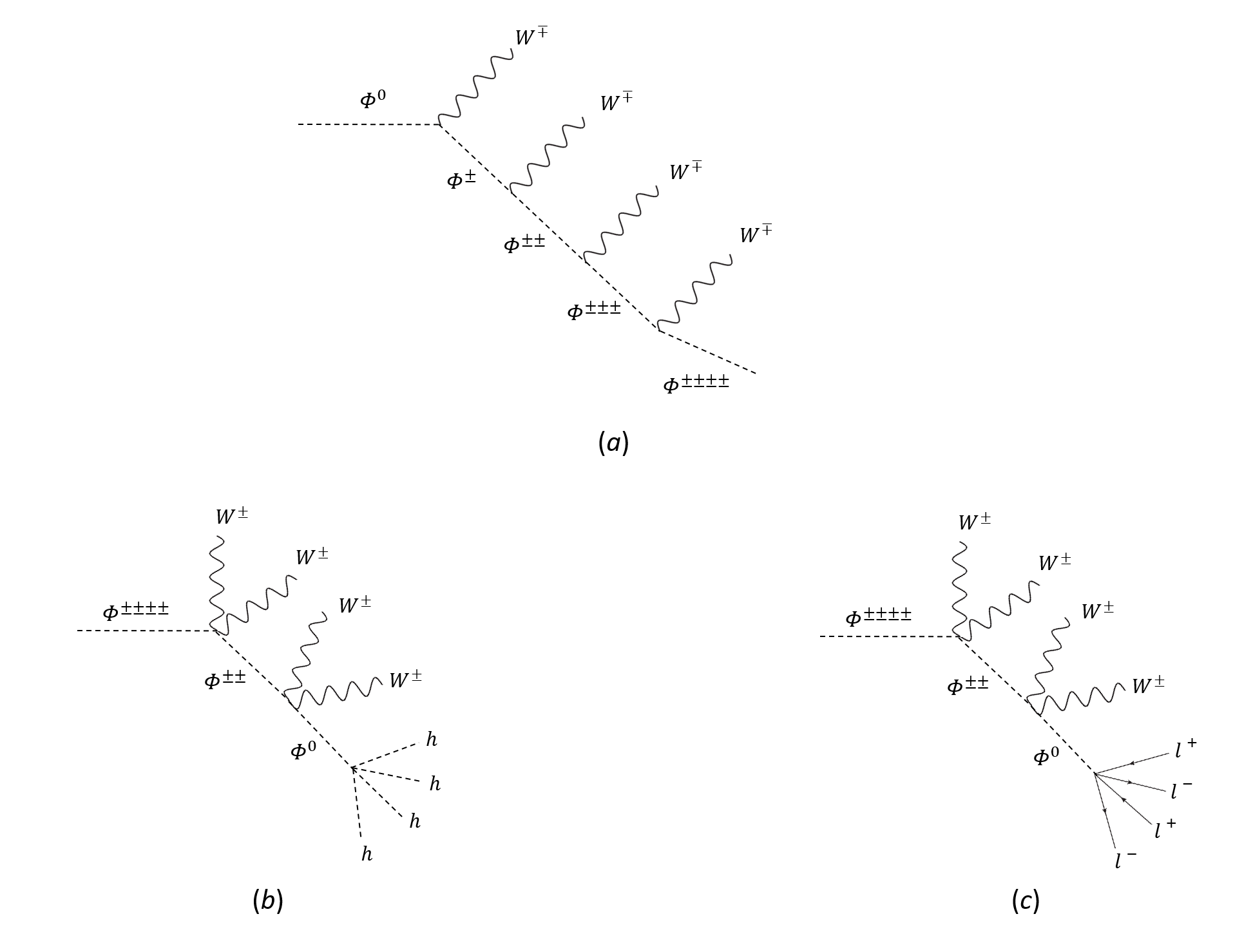}
	\end{center}
	\caption{Decay chains of the scalar particles when the high-dimensional operators  are introduced in the Type-A model with $J=Y=2$.}\label{decay2}
\end{figure}
As illustrated in the plots $(b)$ and $(c)$ of Fig.~\ref{decay2}, the above two operators would cause the lightest charged scalar $\Phi^{\pm \pm \pm \pm}$ to decay into the final states with $4W+4\ell$ or $4W+4h$, in which $\ell$ and $h$ denote the SM leptons and the Higgs scalar, respectively. However, no matter which mechanism breaks the $Z_2$ symmetry, the decays of $\Phi^{\pm \pm \pm \pm}$ would be greatly suppressed by either the small scalar VEV $v_\Phi$ or the high multiplicity of the produced particles in the final states. Hence, a simple estimation shows that the lifetime for a four-charged scalar of ${\cal O}(\rm TeV)$ should be of order of several seconds. This means that this particle should be effectively stable at the LHC, which would only behave as a highly-charged state leaving tracks in the detectors~\cite{ATLAS:2022cob,CMS:2016kce,CMS:2012aza} (also see {\it e.g.} Ref.~\cite{Fairbairn:2006gg} for a theoretical review and references therein). Since the detailed discussion of the LHC signatures of the high-dimensional scalar multiplet are beyond the scope of the present work, we shall leave it in future studies.

\appendix
\section{The Lagrangian and Feynman Rules for EW Gauge Couplings}
\label{app1}
\begin{figure}[ht]
	\begin{center}
		\includegraphics[width=0.9 \linewidth]{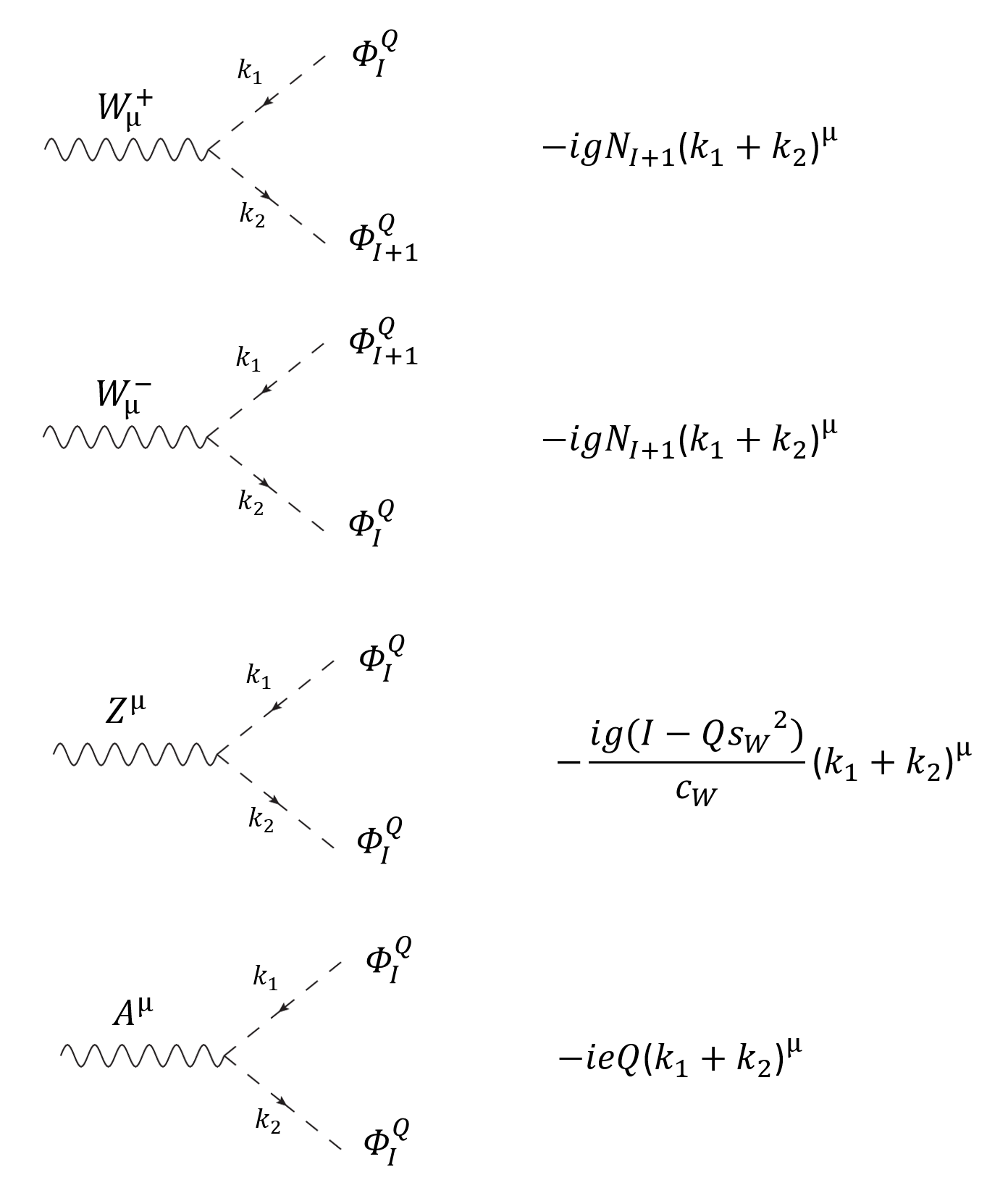}
	\end{center}
	\caption{The three-point EW gauge couplings for components in a multiplet.}\label{3-point-vertex}
\end{figure}

\begin{figure}[ht]
	\begin{center}
		\includegraphics[width=0.9 \linewidth]{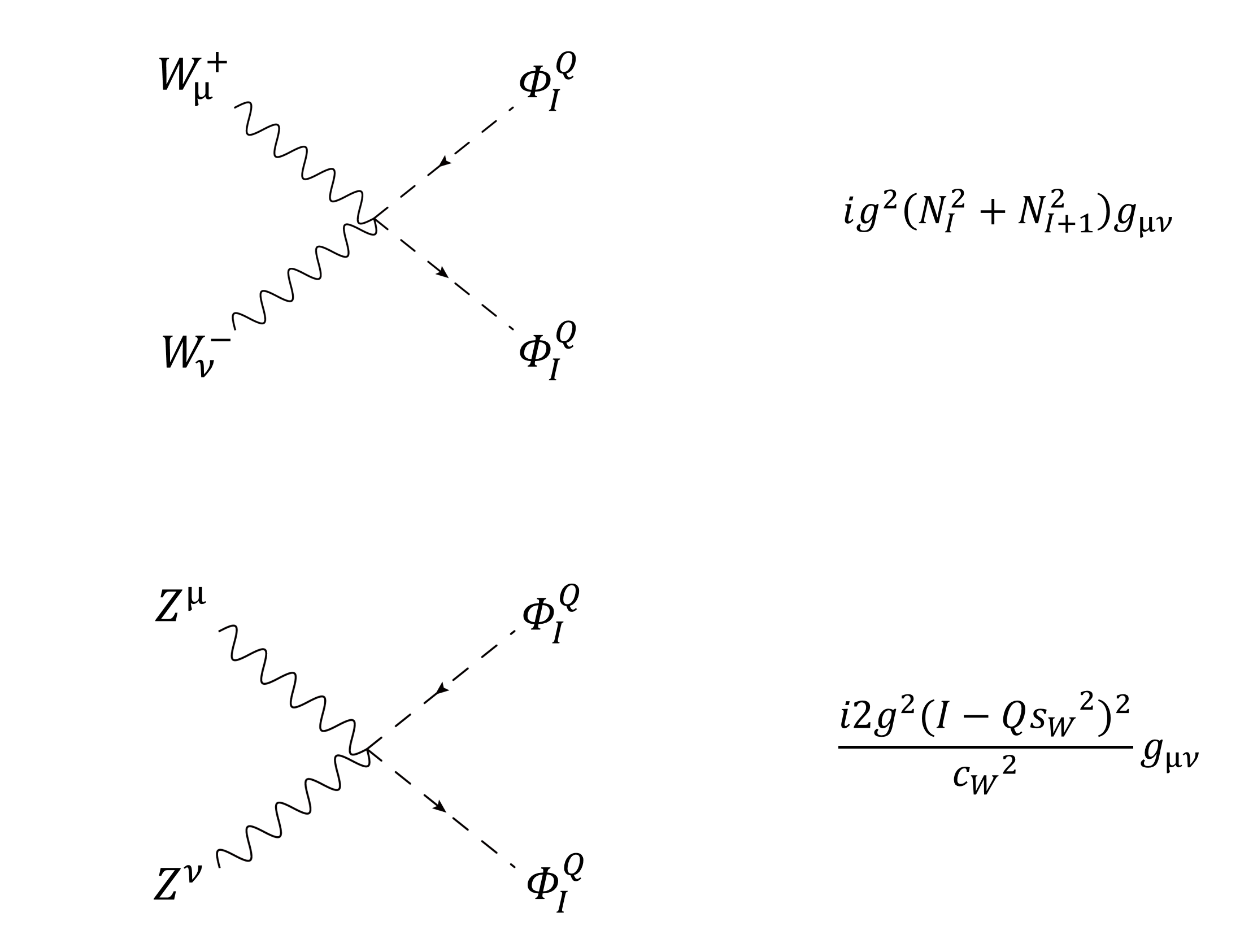}
	\end{center}
	\caption{The four-point EW gauge couplings for components in a multiplet.}\label{4-point-vertex}
\end{figure}

In this Appendix we explicitly write down the EW gauge couplings of a general scalar multiplet in the Lagrangian, which are useful in our calculation of the EW oblique parameters, $T$ and $S$. Firstly, the weak neutral current part can be written as 
\begin{equation}
	\begin{aligned}
		\mathcal{L}_{Z\Phi}&
		=\sum_{I=-J}^{J}\partial_{\mu}\left(\varPhi_{I}^{Q}\right)^{*}\partial^{\mu}\varPhi_{I}^{Q}\\
		&+\sum_{I=-J}^{J}i\frac{g}{c_{W}}\left(I-Qs_{W}^{2}\right)Z_{\mu}\left[\partial_{\mu}\left(\varPhi_{I}^{Q}\right)^{*}\varPhi_{I}^{Q}-\partial^{\mu}\varPhi_{I}^{Q}\left(\varPhi_{I}^{Q}\right)^{*}\right]\\
		&+\sum_{I=-J}^{J}\frac{g^{2}}{c_{W}^{2}}\left(I-Qs_{W}^{2}\right)^{2}Z_{\mu}Z^{\mu}\varPhi_{I}^{Q}\left(\varPhi_{I}^{Q}\right)^{*}\,,
	\end{aligned}
\end{equation}
while the weak charged current part is given by
\begin{equation}
	\begin{aligned}
		\mathcal{L}_{W\Phi}=
		&\sum_{I=-J}^{J}\partial_{\mu}\left(\varPhi_{I}^{Q}\right)^{*}\partial^{\mu}\varPhi_{I}^{Q}\\
		+&\sum_{I=-J}^{J}igW_{\mu}^{+}\left[N_{I}\varPhi_{I-1}^{Q}\partial_{\mu}\left(\varPhi_{I}^{Q}\right)^{*}-N_{I+1}\left(\varPhi_{I+1}^{Q}\right)^{*}\partial^{\mu}\varPhi_{I}^{Q}\right]\\
		+&\sum_{I=-J}^{J}igW_{\mu}^{-}\left[N_{I+1}\varPhi_{I+1}^{Q}\partial_{\mu}\left(\varPhi_{I}^{Q}\right)^{*}-N_{I}\left(\varPhi_{I-1}^{Q}\right)^{*}\partial_{\mu}\varPhi_{I}^{Q}\right]\\
		+&\sum_{I=-J}^{J}g^{2}N_{I+1}^{2}W_{\mu}^{+}W_{\mu}^{-}\left(\varPhi_{I+1}^{Q}\right)^{*}\varPhi_{I+1}^{Q}\\
		+&\sum_{I=-J}^{J}g^{2}N_{I}^{2}W_{\mu}^{+}W_{\mu}^{-}\left(\varPhi_{I-1}^{Q}\right)^{*}\varPhi_{I-1}^{Q}\,,
	\end{aligned}
\end{equation}
where the coefficient $N_I$ is defined below Eq.~(\ref{TMinus}) in Sec.~\ref{sec2.1}. 
Finally, the interaction of scalar components with photons is written as
\begin{equation}
	\begin{aligned}
		\mathcal{L}_{A\Phi}&
		=\sum_{I=-J}^{J}\partial_{\mu}\left(\varPhi_{I}^{Q}\right)^{*}\partial^{\mu}\varPhi_{I}^{Q}\\
		&+\sum_{I=-J}^{J}ieQA_{\mu}\left[\varPhi_{I}^{Q}\partial_{\mu}\left(\varPhi_{I}^{Q}\right)^{*}-\left(\varPhi_{I}^{Q}\right)^{*}\partial_{\mu}\varPhi_{I}^{Q}\right]\,.
	\end{aligned}
\end{equation}
We can derive the Feynman rules for these EW gauge couplings
from the Lagrangian, in which the three-point vertices are shown in Fig.~\ref{3-point-vertex} and the four-point ones in Fig.~\ref{4-point-vertex}.

\section{Possible Potential Terms: $O_6$ and $O_7$}
\label{app2}
In this appendix, we show that the possible potential terms given in Eq.~(\ref{O67}) can be written as the linear combinations of other terms already existing in Eq.~(\ref{potential}). We begin our discussion by expanding the operator $O_{6}$ in terms of scalar components in one multiplet as in Eq.(\ref{rpe}):
\begin{eqnarray}
	O_6 &=&\lambda_{6}\left(H^{\dagger}\Phi_{JY}\right)\left(\Phi_{JY}^{\dagger}H\right) \nonumber\\
	&=&\lambda_{6}\sum_{I=-J}^{J}C_{2J}^{J-I}\left[\left(H^{*}\right)^{a}\left(\varPhi_{I}\right)_{abc...m}\right]\left[\left(\varPhi_{I}^{*}\right)^{abc...m}H_{a}\right] \nonumber\\
	&=&\lambda_{6}\sum_{I=-J}^{J}C_{2J-1}^{J-I-1}\left[\left(H^{*}\right)^{0}\left(\varPhi_{I}\right)_{0}+\left(H^{*}\right)^{1}\left(\varPhi_{I}\right)_{1}\right]_{bc...m}\left[\left(\varPhi_{I}^{*}\right)^{0}H_{0}+\left(\varPhi_{I}^{*}\right)^{1}H_{1}\right]^{bc...m} \nonumber\\
	&=&\lambda_{6}\frac{\left(v+h\right)^{2}}{2}\sum_{I=-J}^{J}C_{2J-1}^{J-I-1}\left(\varPhi_{I}\right)_{0bc...m}\left(\varPhi_{I}^{*}\right)^{0bc...m} \nonumber\\
	&=&\lambda_{6}\frac{\left(v+h\right)^{2}}{2}\frac{C_{2J-1}^{J-I-1}}{C_{2J}^{J-I  }}\varPhi_{I}^{Q}\left(\varPhi_{I}^{Q}\right)^{*} \nonumber\\
	&=&\lambda_{6}\frac{\left(v+h\right)^{2}}{4}\sum_{I=-J}^{J}\varPhi_{I}^{Q}\left(\varPhi_{I}^{Q}\right)^{*}-\frac{\lambda_{6}}{J}\frac{\left(v+h\right)^{2}}{4}\sum_{I=-J}^{J}I\varPhi_{I}^{Q}\left(\varPhi_{I}^{Q}\right)^{*},\label{666}
\end{eqnarray}
where $\left(H^{*}\right)^{1}=H_{1}=0$ and $\left(H^{*}\right)^{0}=H_{0}=\left(v+h\right)/\sqrt{2}$ in the unitary gauge. Also, we have used another notation to represent the components in a multiplet
 \begin{eqnarray}
 \left(\varPhi_{I}\right)_{abc...m}=\frac{1}{\sqrt{C_{2J}^{J-I}}}\varPhi_{I}^{Q}\,, 
\end{eqnarray} 
in which the lowercase Latin indices are those for the $SU(2)_L$ fundamental representation and $C^I_J \equiv J!/[I! (J-I)!]$ is the combinatorial number. According to Eq.~(\ref{666}), the potential term $O_{6}$ can be regarded as a linear combination of terms proportional to $\lambda_{3}$ and $\lambda_{4}$. Therefore, $O_{6}$ is not an independent operator. Similarly, we can prove that the term $O_{7}$ does not represent a new interaction term either.

\section{Feynman Integrals and Functions}
\label{app3}
When computing the oblique parameters $T$ and $S$ from scalar components, we shall encounter the following loop integrals. Since all these integrals are UV divergent, we shall apply the dimensional regularization to obtain the analytic expressions.  For the diagrams $(a_{1,2})$ of Fig.~\ref{fig1} in which only one internal scalar propagator is involved, we need to compute the loop integral~\cite{Grimus:2007if} as follows
\begin{equation}
	\mu^{4-d}\int\frac{d^{d}k}{\left(2\pi\right)^{d}}\frac{g^{\mu\nu}}{k^{2}-A+i\varepsilon}=\frac{ig^{\mu\nu}}{16\pi^{2}}A\left(div-\ln A\right)\,,
\end{equation}
in which $div$ represents the UV divergent part given by
\begin{equation}
	div\equiv\frac{2}{4-d}-\gamma+1+\ln\left(4\pi\mu^{2}\right)\,,
\end{equation}
with $\gamma$ as the Euler's constant, $d$ the spacetime dimensions, and $\mu$ the sliding mass scale, respectively. On the other hand, the diagrams $(b_{1,2})$ in Fig.~\ref{fig1} can also give rise to contributions to the $T$ parameter, in which the internal momentum integrals can be regularized as 
 \begin{equation}
 	\begin{aligned}
 		&\mu^{4-d}\int\frac{d^{d}k}{\left(2\pi\right)^{d}}\int_{0}^{1}dx\frac{4k^{\mu}k^{\nu}}{\left[k^{2}-Ax-B\left(1-x\right)+i\varepsilon\right]^{2}}\\
 		=&\frac{ig^{\mu\nu}}{16\pi^{2}}\left[A\left(div-\ln A\right)+B\left(div-\ln B\right)+F\left(A,B\right)\right]\,,
 	\end{aligned}
 \end{equation}
 where $x$ is the Feynman parameter, $A$ and $B$ denote the masses squared of the scalars in the loop, and the function $F(A,B)$ is defined by
 \begin{equation}\label{Ffunc}
 	F\left(A,B\right)\equiv\begin{cases}
 		\frac{A+B}{2}-\frac{AB}{A-B}\ln\frac{A}{B}\,, & A\neq B\,,\\
 		0\,, & A=B\,.
 	\end{cases}
 \end{equation}

Moreover, in order to obtain the corrections to the oblique parameter $S$, we shall calculate the following integral for Feynman diagrams in Fig.~\ref{fig2}~\cite{Albergaria:2021dmq}:
\begin{equation}
	\begin{aligned}
		\left.\int_{0}^{1}dx\frac{\partial}{\partial q}\left\{ \frac{4}{d}\mu^{4-d}\int\frac{d^{d}k}{\left(2\pi\right)^{d}}\frac{k^{2}}{\left[k^{2}-D\left(q,A,B,x\right)+i\varepsilon\right]^{2}}\right\} \right|_{q=0}&=i\left[div^{*}+\frac{K\left(A,B\right)}{48\pi^{2}}\right.\\
		&+\left.\frac{\ln A+\ln B}{96\pi^{2}}\right]\,,
	\end{aligned}
\end{equation}
where 
\begin{equation}
	D(q,A,B,x)\equiv q^{2}x(x-1)+Ax+B(1-x)\,,
\end{equation}
with $q$ labeling the external four-momentum of gauge bosons,
\begin{align}\label{K}
	K\left(A,B\right)\equiv\begin{cases}
		-\frac{5}{6}+\frac{2AB}{\left(A-B\right)^{2}}+\frac{A^{3}+B^{3}-3AB\left(A+B\right)}{2\left(A-B\right)^{3}}\ln\frac{A}{B}\,, & A\neq B\,,\\
		0\,, & A=B,
	\end{cases}
\end{align}
and 
\begin{equation}
	div^{*}=\frac{-1}{48\pi^{2}}\left[\frac{2}{4-d}-\gamma+\ln\left(4\pi\mu^{2}\right)\right]\,.
\end{equation}

\section{Calculation Details for the Scalar Multiplet Contributions to $T$ and $S$}
\label{app4}
In this appendix, we rederive the analytic expressions for the one-loop contributions to the oblique parameters $T$ and $S$ from a general scalar multiplet. As mentioned in Sec.\ref{sec2.2}, the expression of $T$ is given by
\begin{equation}\label{TWW}
	T\equiv\frac{1}{\alpha m_{Z}^{2}}\left[\frac{A_{WW}\left(0\right)}{c_{W}^{2}}-A_{ZZ}\left(0\right)\right],
\end{equation}
where $A_{VV^\prime}(q)$ is defined in terms of the vacuum polarization $i\Pi_{VV^\prime}^{\mu\nu}(q)$ as in Eq.~(\ref{DefAA}).
It turns out that the vacuum polarization of the $W$ boson is given by
\begin{equation}
	\begin{aligned}
		i\Pi_{WW}^{\mu\nu}\left(0\right)&=\sum_{I=-J}^{J}-g^{2}\left(N_{I}^{2}+N_{I+1}^{2}\right)\mu^{4-d}\int\frac{d^{d}k}{\left(2\pi\right)^{d}}\frac{g^{\mu\nu}}{k^{2}-m_{\varPhi_{I}^{Q}}^{2}+i\varepsilon}\\
		&+\sum_{I=-J}^{J-1}g^{2}N_{I+1}^{2}\mu^{4-d}\int\frac{d^{d}k}{\left(2\pi\right)^{d}}\int_{0}^{1}dx\frac{4k^{\mu}k^{\nu}}{\left[k^{2}-m_{\varPhi_{I}^{Q}}^{2}x-m_{\varPhi_{I-1}^{Q}}^{2}\left(1-x\right)+i\varepsilon\right]^{2}}\\
		&=ig^{\mu\nu}\sum_{I=-J}^{J}\frac{-g^{2}\left(N_{I}^{2}+N_{I+1}^{2}\right)}{16\pi^{2}}m_{\varPhi_{I}^{Q}}^{2}\left(div-\ln m_{\varPhi_{I}^{Q}}^{2}\right)\\
		&+ig^{\mu\nu}\sum_{I=-J}^{J-1}\frac{g^{2}N_{I+1}^{2}}{16\pi^{2}}\left[m_{\varPhi_{I}^{Q}}^{2}\left(div-\ln m_{\varPhi_{I}^{Q}}^{2}\right)+m_{\varPhi_{I+1}^{Q}}^{2}\left(div-\ln m_{\varPhi_{I+1}^{Q}}^{2}\right)\right.\\
		&\left.+F\left(m_{\varPhi_{I}^{Q}}^{2},m_{\varPhi_{I+1}^{Q}}^{2}\right)\right]\,,
	\end{aligned}
\end{equation}
so we can extract $A_{WW}$ as follows
\begin{equation}
	\begin{aligned}
		A_{WW}\left(0\right)&=\sum_{I=-J}^{J}\frac{-g^{2}\left(N_{I}^{2}+N_{I+1}^{2}\right)}{16\pi^{2}}m_{\varPhi_{I}^{Q}}^{2}\left(div-\ln m_{\varPhi_{I}^{Q}}^{2}\right)\\
		&+\sum_{I=-J}^{J-1}\frac{g^{2}N_{I+1}^{2}}{16\pi^{2}}\left[m_{\varPhi_{I}^{Q}}^{2}\left(div-\ln m_{\varPhi_{I}^{Q}}^{2}\right)+m_{\varPhi_{I+1}^{Q}}^{2}\left(div-\ln m_{\varPhi_{I+1}^{Q}}^{2}\right)\right.\\
		&\left.+F\left(m_{\varPhi_{I}^{Q}}^{2},m_{\varPhi_{I+1}^{Q}}^{2}\right)\right]\\
		&=\frac{g^{2}}{16\pi^{2}}\sum_{I=-J}^{J-1}N_{I+1}^{2}F\left(m_{\varPhi_{I}^{Q}}^{2},m_{\varPhi_{I+1}^{Q}}^{2}\right).
	\end{aligned}
\end{equation}
The vacuum polarization of the $Z$ boson is given by
\begin{equation}
	\begin{aligned}
		i\Pi_{ZZ}^{\mu\nu}\left(0\right)&=\sum_{I=-J}^{J}-\frac{2g^{2}\left(I-Qs_{W}^{2}\right)^{2}}{c_{W}^{2}}\mu^{4-d}\int\frac{d^{d}k}{\left(2\pi\right)^{d}}\frac{g^{\mu\nu}}{k^{2}-m_{\varPhi_{I}^{Q}}^{2}+i\varepsilon}\\
		&+\sum_{I=-J}^{J}\frac{g^{2}\left(I-Qs_{W}^{2}\right)^{2}}{c_{W}^{2}}\mu^{4-d}\int\frac{d^{d}k}{\left(2\pi\right)^{d}}\int_{0}^{1}dx\frac{4k^{\mu}k^{\nu}}{\left[k^{2}-m_{\varPhi_{I}^{Q}}^{2}+i\varepsilon\right]^{2}}\\
		&=ig^{\mu\nu}\sum_{I=-J}^{J}-\frac{2g^{2}\left(I-Qs_{W}^{2}\right)^{2}}{16\pi^{2}c_{W}^{2}}m_{\varPhi_{I}^{Q}}^{2}\left(div-\ln m_{\varPhi_{I}^{Q}}^{2}\right)\\
		&+ig^{\mu\nu}\sum_{I=-J}^{J}\frac{2g^{2}\left(I-Qs_{W}^{2}\right)^{2}}{16\pi^{2}c_{W}^{2}}m_{\varPhi_{I}^{Q}}^{2}\left(div-\ln m_{\varPhi_{I}^{Q}}^{2}\right)\,,\\
	\end{aligned}
\end{equation}
so we have
\begin{equation}
	\begin{aligned}
		A_{ZZ}\left(0\right)&=\sum_{I=-J}^{J}-\frac{2g^{2}\left(I-Qs_{W}^{2}\right)^{2}}{16\pi^{2}c_{W}^{2}}m_{\varPhi_{I}^{Q}}^{2}\left(div-\ln m_{\varPhi_{I}^{Q}}^{2}\right)\\
		&+\sum_{I=-J}^{J}\frac{2g^{2}\left(I-Qs_{W}^{2}\right)^{2}}{16\pi^{2}c_{W}^{2}}m_{\varPhi_{I}^{Q}}^{2}\left(div-\ln m_{\varPhi_{I}^{Q}}^{2}\right)\\
		&=0\,.
	\end{aligned}
\end{equation}
Therefore, the contribution of scalar multiplet $\Phi_{JY}$ to $T$ is only provided by the $A_{WW}$ part in Eq.~(\ref{TWW})
\begin{equation}
	\begin{aligned}
		T_{\Phi_{JY}}&=\frac{g^{2}}{16\alpha\pi^{2}c_{w}^{2}m_{Z}^{2}}\sum_{I=-J}^{J-1}N_{I+1}^{2}F\left(m_{\varPhi_{I}^{Q}}^{2},m_{\varPhi_{I+1}^{Q}}^{2}\right)\\
		&=\frac{1}{4\pi s_{w}^{2}m_{W}^{2}}\sum_{I=-J}^{J-1}N_{I+1}^{2}F\left(m_{\varPhi_{I}^{Q}}^{2},m_{\varPhi_{I+1}^{Q}}^{2}\right)\,.
	\end{aligned}
\end{equation}

On the other hand, the expression of $S$ is given by
\begin{equation}
	S\equiv\frac{4s_{W}^{2}c_{W}^{2}}{\alpha}\left[A_{ZZ}^{\prime}\left(0\right)-\frac{c_{W}^{2}-s_{W}^{2}}{c_{W}s_{W}}A_{Z\gamma}^{\prime}\left(0\right)-A_{\gamma\gamma}^{\prime}\left(0\right)\right]\,,
\end{equation}
where $A^\prime_{VV^\prime}$ is defined in Eq.~(\ref{DefAA}) as the expansion of $\Pi^{\mu\nu}_{VV^\prime}$ in terms of the external momentum $q$ at the second order.  
Firstly, we calculate the $ZZ$ part:
\begin{equation}
	\begin{aligned}
		&A_{ZZ}^{\prime}\left(0\right)\\
		&=-i\sum_{I=-J}^{J}\frac{g^{2}\left(I-Qs_{W}^{2}\right)^{2}}{c_{W}^{2}}\left.\int_{0}^{1}dx\frac{\partial}{\partial q}\left\{ \frac{4}{d}\mu^{4-d}\int\frac{d^{d}k}{\left(2\pi\right)^{d}}\frac{k^{2}}{\left[k^{2}-D\left(q,m_{\varPhi_{I}^{Q}}^{2},m_{\varPhi_{I}^{Q}}^{2},x\right)+i\varepsilon\right]^{2}}\right\} \right|_{q=0}\\
		&=\sum_{I=-J}^{J}\frac{g^{2}\left(I-Qs_{W}^{2}\right)^{2}}{c_{W}^{2}}\left[div^{*}+\frac{K\left(m_{\varPhi_{I}^{Q}}^{2},m_{\varPhi_{I}^{Q}}^{2}\right)}{48\pi^{2}}+\frac{\ln m_{\varPhi_{I}^{Q}}^{2}}{48\pi^{2}}\right]\\
		&=\sum_{I=-J}^{J}\frac{g^{2}\left(I-Qs_{W}^{2}\right)^{2}}{c_{W}^{2}}\left(div^{*}+\frac{\ln m_{\varPhi_{I}^{Q}}^{2}}{48\pi^{2}}\right)\,,
	\end{aligned}
\end{equation}
where
\begin{equation}
	K\left(m_{\varPhi_{I}^{Q}}^{2},m_{\varPhi_{I}^{Q}}^{2}\right)=0\,,
\end{equation}
in the light of Eq.(\ref{K}).
Then, we calculate the $Z\gamma$ part:
\begin{equation}
	\begin{aligned}
		&A_{Z\gamma}^{\prime}\left(0\right)\\
		&=-i\sum_{I=-J}^{J}\frac{egQ\left(I-Qs_{W}^{2}\right)}{c_{W}}\left.\int_{0}^{1}dx\frac{\partial}{\partial q}\left\{ \frac{4}{d}\mu^{4-d}\int\frac{d^{d}k}{\left(2\pi\right)^{d}}\frac{k^{2}}{\left[k^{2}-D\left(q,m_{\varPhi_{I}^{Q}}^{2},m_{\varPhi_{I}^{Q}}^{2},x\right)+i\varepsilon\right]^{2}}\right\} \right|_{q=0}\\
		&=\sum_{I=-J}^{J}\frac{egQ\left(I-Qs_{W}^{2}\right)}{c_{W}}\left[div^{*}+\frac{K\left(m_{\varPhi_{I}^{Q}}^{2},m_{\varPhi_{I}^{Q}}^{2}\right)}{48\pi^{2}}+\frac{\ln m_{\varPhi_{I}^{Q}}^{2}}{48\pi^{2}}\right]\\
		&=\sum_{I=-J}^{J}\frac{egQ\left(I-Qs_{W}^{2}\right)}{c_{W}}\left(div^{*}+\frac{\ln m_{\varPhi_{I}^{Q}}^{2}}{48\pi^{2}}\right)\,.
	\end{aligned}
\end{equation}
Finally, we calculate the $\gamma\gamma$ part:
\begin{equation}
	\begin{aligned}
		&A_{\gamma\gamma}^{\prime}\left(0\right)\\
		&=-i\sum_{I=-J}^{J}e^{2}Q^{2}\left.\int_{0}^{1}dx\frac{\partial}{\partial q}\left\{ \frac{4}{d}\mu^{4-d}\int\frac{d^{d}k}{\left(2\pi\right)^{d}}\frac{k^{2}}{\left[k^{2}-D\left(q,m_{\varPhi_{I}^{Q}}^{2},m_{\varPhi_{I}^{Q}}^{2},x\right)+i\varepsilon\right]^{2}}\right\} \right|_{q=0}\\
		&=\sum_{I=-J}^{J}e^{2}Q^{2}\left[div^{*}+\frac{K\left(m_{\varPhi_{I}^{Q}}^{2},m_{\varPhi_{I}^{Q}}^{2}\right)}{48\pi^{2}}+\frac{\ln m_{\varPhi_{I}^{Q}}^{2}}{48\pi^{2}}\right]\\
		&=\sum_{I=-J}^{J}e^{2}Q^{2}\left(div^{*}+\frac{\ln m_{\varPhi_{I}^{Q}}^{2}}{48\pi^{2}}\right).
	\end{aligned}
\end{equation}
Therefore, the contribution of a scalar multiplet $\Phi_{JY}$ to $S$ is given by
\begin{equation}\label{SJY}
	\begin{aligned}
		S_{\Phi_{JY}}&=\sum_{I=-J}^{J}\frac{4s_{W}^{2}c_{W}^{2}}{\alpha}\left\{\frac{g^{2}\left[I-\left(I+Y\right)s_{W}^{2}\right]^{2}}{c_{W}^{2}}-e^{2}\left(I+Y\right)^{2}\right.\\
		&\left.-\frac{eg\left(I+Y\right)\left(c_{W}^{2}-s_{W}^{2}\right)\left[I-\left(I+Y\right)s_{W}^{2}\right]}{c_{W}^{2}s_{W}}\right\}\left(div^{*}+\frac{\ln m_{\varPhi_{I}^{Q}}^{2}}{48\pi^{2}}\right)\\
		&=-16\pi Y\sum_{I=-J}^{J}I\left(div^{*}+\frac{\ln m_{\varPhi_{I}^{Q}}^{2}}{48\pi^{2}}\right)\,.
	\end{aligned}
\end{equation}
Since the sum over the isospin third components $I$ vanishes identically, {\it i.e.},
\begin{equation}
	Y\sum_{I=-J}^{J}I=0\,,
\end{equation}
the UV divergences in Eq.~(\ref{SJY}) are cancelled. Thus, the contribution of a scalar multiplet $\Phi_{JY}$ to $S$ can be written as
\begin{equation}
	S_{\Phi_{JY}}=-\frac{Y}{3\pi}\sum_{I=-J}^{J}I\ln m_{\varPhi_{I}^{Q}}^{2}.
\end{equation}
Note that the expressions of $S_{\Phi_{JY}}$ and $T_{\Phi_{JY}}$ are consistent with those given in Ref.~\cite{Lavoura:1993nq}.

\acknowledgments
This work is supported in part by the National Key Research and Development Program of China (Grant No.~2021YFC2203003 and No.~2020YFC2201501 )
and   the National Natural Science Foundation of China (NSFC) (Grant No. 12005254 and No. 12147103).


\end{document}